\documentclass[showpacs,twocolumn,longbibliography]{revtex4-2}
\usepackage{epstopdf}
\usepackage{color}
\usepackage{bm}
\usepackage{graphicx}
\usepackage{amsmath}
\usepackage{amsfonts}
\usepackage{amssymb}
\usepackage{bezier} 
\usepackage{color} 
\usepackage{graphicx,amsmath,mathrsfs}
\usepackage{amssymb}
\usepackage{amsthm,multirow}
\usepackage{bm,bbm}
\usepackage{subfigure,xcolor} 
\usepackage[makeroom]{cancel}
\usepackage{color}
\usepackage{float}
\usepackage{bbold}

\def\nd{^{\vphantom{\dagger}}}

\def\bB{{\bf B}}

\def\bx{{\bf x}}

\def\br{{\bf r}}
\def\bz{{\bf z}}

\def\bs{{\bf s}}

\def\bk{{\bf k}}

\def\cL{{\cal L}}

\def\cP{{\cal P}}

\def\cO{{\cal O}}

\def\cM{{\cal M}}

\def\tc{{\tilde c}}
\def\Im{{\mbox{Im}}}
\def\Re{{\mbox{Re}}}
\def\Tr{{\mbox{Tr}}}

\def\ij{{\langle ij\rangle}}
\def\jk{{\langle jk\rangle}}

\def\ijk{{\langle ijk\rangle}}
\def\be{\begin{equation}}
\def\ee{\end{equation}}
\def\bea{\begin{eqnarray}}
\def\eea{\end{eqnarray}}
\def\sgn{\mbox{\rm sgn}}

\def\lsco{La$_{2-x}$Sr$_x$CuO$_4$}
\def\ybco{YBa$_2$Cu$_3$O$_y$}
\def\tt{{t}}

\begin{document}
%--------------------------------------------------------------------------------
\title{Hall map and breakdown of Fermi liquid theory in the vicinity of a Mott insulator}

\author{Ilia Khait$^1$, Sauri Bhattacharyya$^2$, Abhisek Samanta$^2$ and  Assa Auerbach$^2$}
\affiliation{$^1$Entangled Networks Ltd., M4R 2E4 Toronto, Ontario, Canada,\\ 
$^2$Physics Department, Technion, 32000 Haifa, Israel.}

\date{\today }
\begin{abstract}   
The Hall coefficient exhibits anomalous behavior in lightly doped Mott insulators. For strongly interacting electrons its computation has been challenged
by  analytical  and numerical obstacles. We  calculate the leading contributions in the recently derived thermodynamic formula for the Hall coefficient.
We obtain its doping and temperature dependence for the square lattice tJ-model at high temperatures.  The second order corrections are evaluated  to be negligible.  Quantum Monte Carlo sampling extends our results to lower temperatures.  We find a divergence of the Hall coefficient toward the Mott limit
and a sign reversal relative to Boltzmann equation's weak scattering prediction.  The Hall current near the Mott phase is carried by a low density of spin-entangled vacancies, which should constitute the Cooper pairs in any superconducting phase at lower temperatures. 
 \end{abstract}

%\pacs{72.10.Bg,72.15.-v, 72.15.Gd}

\maketitle

\section{Introduction}
Doped Mott insulators~\cite{Mott} have risen to prominence at the advent of high temperature superconductivity in cuprates~\cite{RVB}. 
Microscopic understanding of these superconductors requires identifying their constituent charge carriers.
Their high temperature  resistivities have been characterized as  ``strange metals'', whose strong scattering is inconsistent with Fermi liquid quasiparticles~\cite{badmetals,PP,Hussey-Review}.   
Moreover, the Hall coefficient $R_{\rm H}$ of e.g. underdoped  \lsco~\cite{Takagi-LSCO,Ando-RH} and \ybco~\cite{YBCO}, appears to diverge toward half filling. This fundamentally contradicts
Fermi liquid based transport theory~\cite{Comm-FLT}.  

\begin{figure}[h!]
\begin{center}
\includegraphics[width=8cm,angle=0]{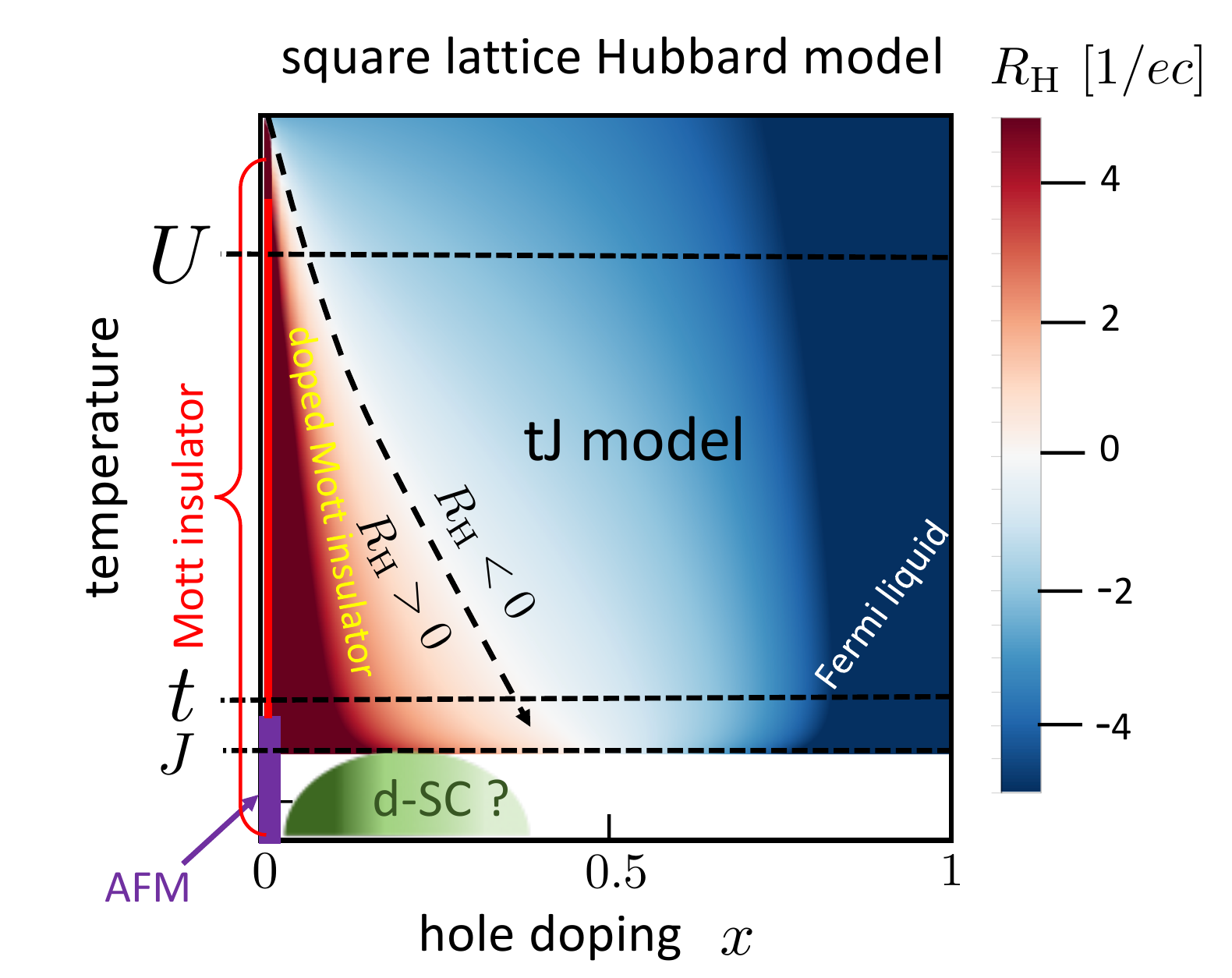}
\caption{
Hall coefficient $R_{\rm H}$ of the Hubbard Model with interaction $U$, and hopping $t$ on the square lattice (with unit lattice constant), as a function of hole doping per site $x$.
We use the corresponding tJ-model at temperatures below the interaction scale $U$.  
In the low doping regime  the Hall sign is reversed relative to the non-interacting model (see Fig.~\ref{fig:HallMap-SL}), and  diverges.   
Antiferromagnetic (AFM) order sets in at $x=0$ below the magnetic energy scale $J$.   A possible $d$-wave superconducting (d-SC)  phase at low doping must  inherit  its charge carriers from the anomalous metallic state.}
 \label{fig:HallMap-U}
\end{center}
\end{figure}
\begin{figure}[h!]
\begin{center}
\includegraphics[width=8cm,angle=0]{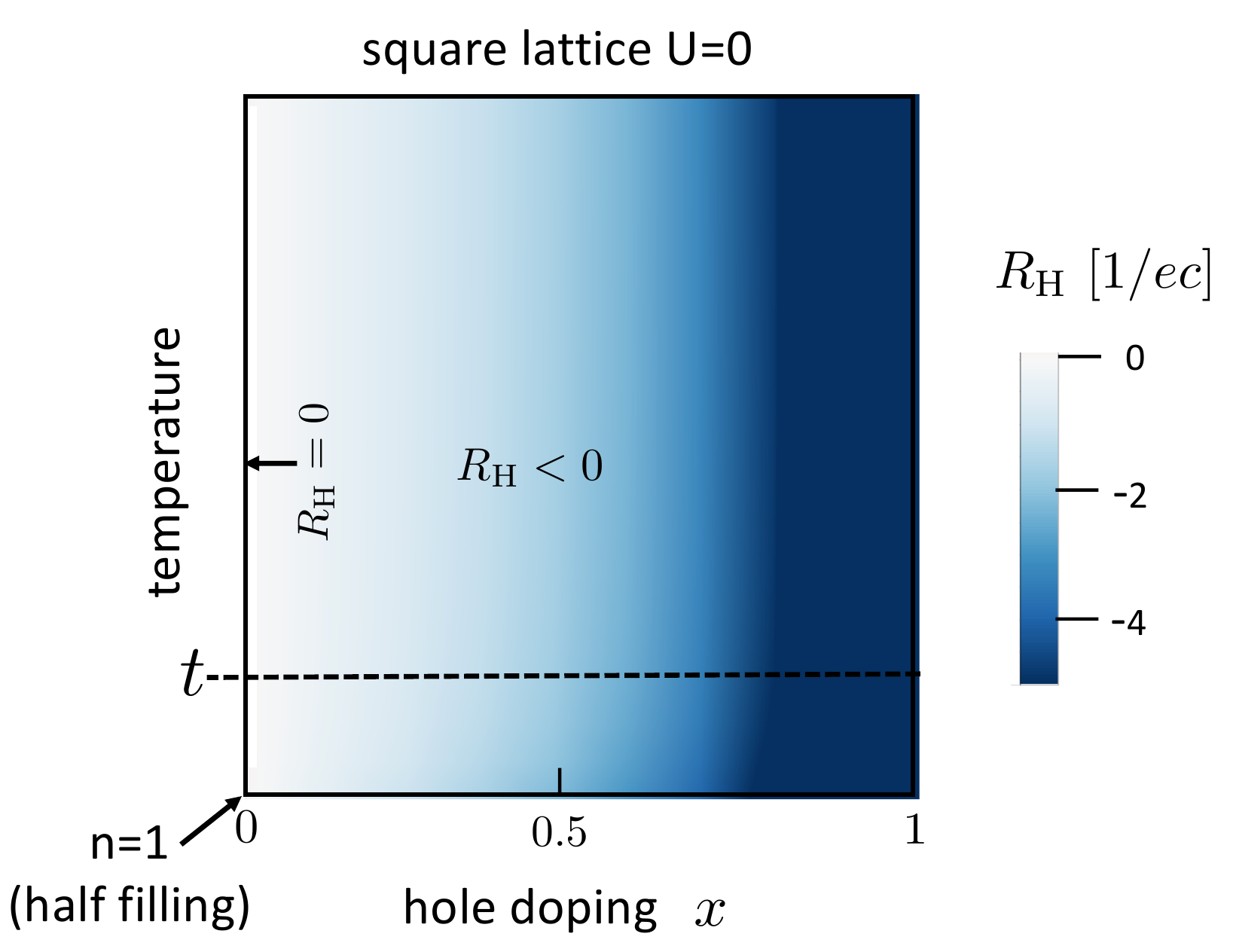}
\caption{
Hall coefficient  of non-interacting electrons on a square lattice evaluated by Boltzmann theory, Eq.~(\ref{RH-Boltz}) for the a  wavevector-independent scattering time.  Note that $R_{\rm H}\!<\!0$ for $0\!<\!x\!<\!1$,  and  {\em no divergence} except at the empty band limit $x\to 1$.  }
 \label{fig:HallMap-SL}
\end{center}
\end{figure}

A minimal model for doped Mott insulators is the
strongly interacting ($U\!\gg\!t$) Hubbard model (HM)~\cite{Hubbard,Kivelson-Hubbard}. At
half filling (doping $x\!=\!0$) interactions open an insulating charge gap.  At temperatures $T<U$~\cite{IEQM},  its low frequency correlations are described by  the tJ-model (tJM)~\cite{tJ-Spalek,IEQM}. Variational studies of the square lattice HM and tJM, found $d$-wave superconductivity and charge  ordering, depending on the details of the hopping terms~\cite{DMRG,Troyer,Sorella}.  Proxies for the Hall coefficient have been calculated~\cite{Prelovsek,HallDMFT,Dev-PR,Khomskii}, but their error estimation remained a challenge.
Quantum Monte Carlo (QMC) computations require an analytic continuation which is challenging in the DC limit~\cite{Comm:QMC-AC}.  

A new  thermodynamic formula for $R_{\rm H}$ includes a generally easy to compute  ratio of susceptibilities $R_{\rm H}^{(0)}$  and a more complicated (but hopefully small) correction term $R^{\rm corr}_{\rm H}$~\cite{EMTPRL,EMT,Abhisek}. 
$R_{\rm H}^{(0)}$ was previously computed  for the HM~\cite{Dev-Rxy} by QMC.  However for the HM,  $R^{\rm corr}_{\rm H}$ increases with the interaction parameter $U/t$ 
and cannot be ignored, especially in the  intermediate temperature (IT) regime defined by
$t/2\le T \le U$.

This paper calculates the $R_{\rm H}^{(0)}$ for $U/t\gg 1$ by replacing the HM by the effective tJM. 
The doping dependence is obtained analytically by a high temperature expansion, and the lower temperatures in the IT regime by  QMC sampling.
Second order corrections of $R^{\rm corr}_{\rm H}$  for the tJM are calculated. They are found to  be negligible relative to $R_{\rm H}^{(0)}$, and
higher order corrections are estimated to be even smaller due to diminishing operator overlaps.  

The temperature-doping Hall map  is depicted in Fig.~\ref{fig:HallMap-U}. It exhibits a substantial region of  positive Hall coefficient, which diverges as $R_{\rm H} \!\propto\! 1/x$ at temperatures lower than the $U$.
The effects of the strong interactions are apparent by contrasting Fig.~\ref{fig:HallMap-U}  with Fig.~\ref{fig:HallMap-SL}, which plots Boltzmann's  non-interacting result for the same  square lattice with   dispersion  $\epsilon_\bk=-2t\left(\cos(k_x)+\cos(k_y)\right)$ and isotropic scattering~\cite{Ziman}:
\be 
R_{\rm H}^{\rm Boltz}=  {   \int {d^2k\over (2\pi^2)}\left(-{df\over d\epsilon}\right)  \left({\partial \epsilon\over\partial k_y}\right)^2 {\partial^2 \epsilon\over (\partial k_x)^2} 
\over 2 \left(\int {d^2k\over (2\pi^2)} \left(-{df\over d\epsilon}\right)   \left({\partial \epsilon\over\partial k_x}\right)^2 \right)^2},
\label{RH-Boltz}
\ee
where $f(\epsilon)$ is Fermi function.
$R_{\rm H}^{\rm Boltz}$  yields a negative Hall coefficient at all hole dopings. Even with the addition of next nearest neighbor hopping, which is often included to fit
the cuprates'  Fermi  arcs, $R_{\rm H}^{\rm Boltz}$  would not be expected to diverge toward half filling.

At  very high temperatures $T\gg U$, the effects of interactions in the HM are suppressed and  $R_{\rm H}(x,T)\to R^{\rm Boltz}_{\rm H}(x,T)$.   
This regime may  be accessed by cold atom simulations of the HM~\cite{Bakr}.

The Hall anomalies near the Mott phase are a consequence of  the effective Gutzwiller projection (GP) in the IT regime~\cite{Gutzwiller}.
The dynamical longitudinal conductivity is highly suppressed relative to the non-interacting limit. The
sign reversal of $R_{\rm H}$ is due to interaction-driven density and spin operators which contribute to the commutators between GP currents and Hamiltonian.   
Thus, we learn that the currents are carried by a low density of spin-entangled {\em positive} vacancies moving in a paramagnetic environment.  
At lower temperatures,  pairs of these projected hole carriers would form the expected~\cite{Sorella} superconducting condensate, as proposed by Anderson~\cite{RVB}.

The paper is organized as follows. The thermodynamic Hall coefficient formula, the HM and the tJM are formally introduced.  
The analytical high temperature expansion of the relevant susceptibilities for the   tJM  is presented.  
Numerical extension to lower temperatures by
QMC simulations is displayed.  
The  correction term is estimated, by evaluation of the second order contribution, and arguments for rapidly diminishing higher orders.  
Previous calculations of $R_{\rm H}$ using different methods are compared to our results.  We conclude by  summarizing the effects of strong Hubbard interactions  on the charge carriers, 
and their implication on lower temperature transport in the expected superconducting flux flow regime~\cite{SciPost}.

\section{The Thermodynamic Hall Coefficient formula}
The  Hall coefficient $R_{\rm H}$, for a magnetic field $\bB=B\hat{\bz}$, is defined by elements of the conductivity tensor $\sigma_{\alpha\beta}$, 
\be
R_{\rm H}  \equiv  {d\sigma_{yx}\over dB} \sigma^{-2}_{xx}\Big|_{B\!=\!0} 
\ee
which  can be expressed by the thermodynamic formula $R_{\rm H}= R_{\rm H}^{(0)} +R^{\rm corr}_{\rm H} $~\cite{EMT}.  Both terms in $R_{\rm H}$ are composed of  thermodynamic averages which are amenable to expansion in powers of inverse temperature $\beta$, and  
QMC {\em without} analytic continuation~\cite{Comm:QMC-AC}. 

As derived in Refs.~\cite{EMTPRL,EMT}, the first term $R_{\rm H}^{(0)}$ is a ratio of the current-magnetization-current (CMC) susceptibility and the conductivity sum rule (CSR) squared:
\bea
&&R_{\rm H}^{(0)}   =
{\chi^{ }_{\rm cmc} \over \chi_{\rm csr}^2 },\nonumber\\
&& \chi_{\rm cmc}  \!=\!     -2  \Re  \langle \big[ P^y ,[ M, j^x]\big]\rangle , ~ \chi_{\rm csr}\!=\!  \Im  \langle \left[ P^x ,  j^x  \right] \rangle , 
\label{Chis}
\eea
where $\langle \cdot \rangle$ is the thermal expectation value. The operators $P^\alpha, j^\alpha$ are the  polarization and current in the $\alpha$ direction, and $M=-{\partial H\over \partial B}$ is the $z$-magnetization.           

The correction term $R_{\rm H}^{\rm corr}$ is an infinite  convergent sum which is defined in  Appendix \ref{App:Corr}. Since $R_{\rm H}^{\rm corr}$ is much harder to calculate, the formula is useful if it can be estimated to be  negligibly small. 
For the tJM at high temperatures, we provide such an estimate  in Section \ref{sec:Corr}, by computing its leading orders as detailed in Appendix \ref{App:Corr}. 

\section{Hubbard  and ${\rm tJ}$ Models}
\label{sec:models}
The square lattice  HM (with units of $\hbar\!=\!1$) is,
\be
H^{\rm HM}=-t\sum_{\ij,s=\uparrow,\downarrow} (c^\dagger_{is}c^{\nd}_{js} +c^\dagger_{js}c^{\nd}_{is}) + U\sum_i n_{i\uparrow}n_{i\downarrow} ,
\label{HM}
\ee
where $c^\dagger_{is}$ creates an electron on site $i$ with spin $s$. $\ij$ are nearest neighbor bonds, and $n_{is}= c^\dagger_{is} c^{}_{is}$.
The Hall coefficient of the non-interacting  model on the square lattice is negative (electron-like) without divergences for all  $0\le x<1$,  as shown in Fig.~\ref{fig:HallMap-SL}.

For $U/t \gg 1$, the low energy subspace is defined by the GP operator 
$\cP_{\rm GP} 
\!=\!\prod_{i}(1- n_{i\uparrow}n_{i\downarrow})$.  The GP electron creation, hole density and spin operators are,
 \bea
&&\tc^\dagger_{is}  \equiv  c^\dagger_{is} (1-n_{i,-s}) ,\quad
n^h_i \equiv 1- \sum_s \tc^\dagger_{is} \tc^{\nd}_{is}    , \nonumber\\
&& s^\alpha_i (1-n^h_i)   \equiv    {1\over 2} \sum_{ss'} \tc^\dagger_{is}\sigma^\alpha \tc^{\nd}_{is'} .
\eea
The electric polarization is defined by $P^\alpha = - e\sum_i x_i^\alpha n^h_i $, where $e$ is the negative electron charge, and $\bx_i$ is the position of site $i$.

The GP hopping terms are,
\bea
&&{K}^{\pm}_{ij}  \equiv    \sum_s \tc^\dagger_{is} \tc^{\nd}_{j s} \pm \tc^\dagger_{js} \tc^{\nd}_{i s}, \nonumber\\
&& {\bf \Sigma}^{\pm}_{ij}   \equiv   \sum_{ss'} \tc^\dagger_{is} \vec{\sigma}_{ss'} \tc^{\nd}_{js'}  \pm  \tc^\dagger_{js}\vec{\sigma}_{ss'} \tc^{\nd}_{is'}
 \label{K-def}
\eea
where $K^+$ ($K^-$) describes the bond  kinetic energy (current). 

The adjacent-bond  commutators for $a,a'=\pm$ are  
%\bea
%[K^{a}_{12},K^{a'}_{23}] & = &  \left(K^{-aa'}_{13} \left( {1+  n^h_{2} \over 2}\right) \right. \nonumber\\ 
%&&+ \left.{\bf \Sigma}^{-aa'}_{13} \cdot   {\bf s}_{2}(1-n^h_2)\right) .  
%\label{comm}
%\eea
\bea
\!\!\!\!\!\! [K^{a}_{12},K^{a'}_{23}] & = &  K^{-aa'}_{13} \left( {1+  n^h_{2} \over 2}\right) 
+ {\bf \Sigma}^{-aa'}_{13} \cdot   {\bf s}_{2}(1-n^h_2) .  
\label{comm}
\eea
Note that these involve a hole density or spin operator on site $2$, which does not appear for a commutator of adjacent unprojected hopping operators~\footnote{Can be easily verified with the multiplication identities in Appendix~\ref{App:Corr}  Table~\ref{table:opmult}.}.
These extra operators are responsible for the sign reversal of $R_{\rm H}$ in the tJM at finite values of doping.

The tJM as derived from Eq.~(\ref{HM})~\cite{tJ-Spalek,IEQM}  can be expressed using Eq.~(\ref{K-def}),
\bea
H^{\rm tJM} &=&  H^t+H^J  +\cO(t^3/U^2)\nonumber\\
H^t  &=&  -\tt\sum_{\ij }  K^+_{ij} ,\nonumber\\
H^{J}   &=& - {J \over 4} \sum_{\ij \jk}   \left(   K^+_{ik} - 2{\bf \Sigma}^+_{ik}\cdot \bs_j\right)(1\!-\!n^h_j).
\label{tJM}
\eea
In the  IT  regime, $R_{\rm H}$ is largely determined by the hole hopping terms of $H^t$.  
$H^J$ scales with the superexchange energy $J=4t^2/U\! \ll \! t$, and includes spin interactions in its diagonal term, $i\!=\!k$, and (often neglected) next neighbor hopping terms $i\!\ne \!k$.
The latter terms actually dominate to the Hall coefficient at very high temperatures $T\sim U$.
 For $H^t$, the current and magnetization are
 \bea
&& j^\alpha_{ij}  = - i {e \over \hbar a}\tt K^-_{ij;\alpha}    , \nonumber\\
&&M = { 1 \over 2c} \sum_\ij  x_i j^y_{ ij}- y_i j^x_{ ij}   .
\label{tModel-op}
\eea
Here, $ij;\alpha$, denotes a directed bond in the $\alpha$ direction, and $c$ is the speed of light. 
The additional contributions of $H^J$ to the current and magnetization are discussed in Section \ref{sec:VHT}.

\begin{figure}[h!]
\begin{center}
\includegraphics[width=7cm,angle=0]{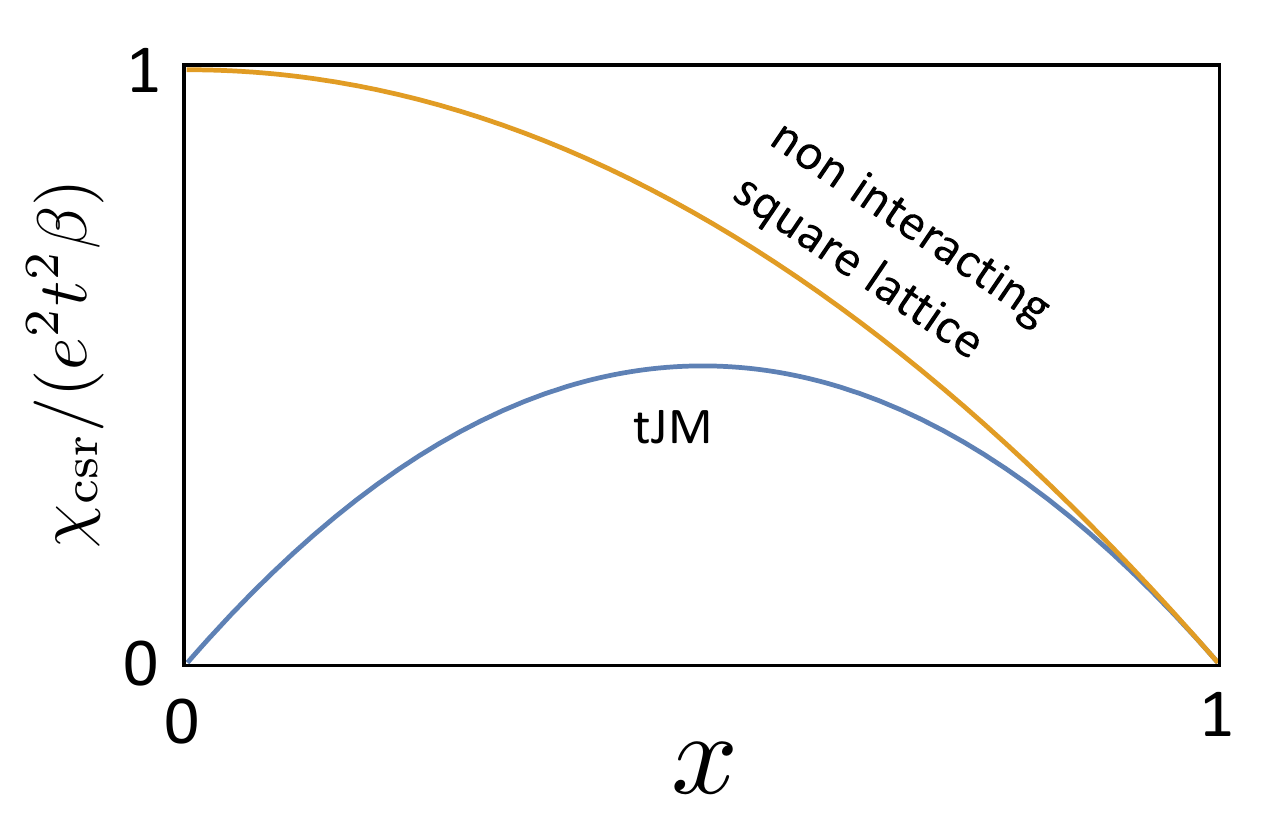}
\caption{Doping dependent CSRs at intermediate temperatures.
The suppression of the tJM CSR relative to the non-interacting square lattice CSR,  affects a large region of doping.
Vanishing of the tJM CSR  at $x\to 0$ leads to
 the anomalous divergence of $R_{\rm H}$ toward the Mott phase, and the diverging resistivity slope.}
 \label{fig:CSR}
\end{center}
\end{figure}

\section{Hall coefficient of the \lowercase{t}JM}
In the IT regime,   $R_{\rm H}$ of the tJM (\ref{tJM})  is dominated by $H^t$.
For the CSR, the doping dependence of the two leading powers of inverse temperature $\beta$ were previously calculated by Jaklic~\cite{Jak} and Perepelitsky~\cite{Perepel}. The calculation is reviewed in Appendix \ref{App:highTexp}, and yields
\bea
\!\!\!\! \chi^t_{\rm csr}&=&2\beta e^2\tt^2 x(1-x)\nonumber\\
&+&{ \beta^3 e^2  \tt^4   \over 6}  x(1-x) (-9 + 2 x + x^2 ) + \cO(\beta^5\tt^6).
\label{CSR-hiT}
\eea
As depicted in  Fig.~\ref{fig:CSR}, the CSR (and therefore the whole dynamical longitudinal  conductivity)  of the tJM vanishes toward $x\to 0$, and is suppressed in a large region of doping.
In contrast, the non-interacting  CSR is maximized at half filling, as expected for a large Fermi surface.
 
The CMC of $H^t$  is evaluated up to order $(\beta \tt)^4 $ in Appendix \ref{App:highTexp},
\bea
\chi^t_{\rm cmc} &=&   - {\beta^2\tt^4 e^3  \over 2c }   x(1-x)  (-5+10x  + 3x^2 )\nonumber\\
&+&   {\beta^4\tt^6 e^3\over 16 c} x(1-x) (45 - 136 x + 50 x^2 +48 x^3 - 71 x^4)  \nonumber\\
&&~~~~~~~~~~~~~~~~~~~~~~~+ \cO(\beta^6\tt^8).
\label{CMC-hiT}
\eea

\begin{figure}[h!]
\begin{center}
\includegraphics[width=8cm,angle=0]{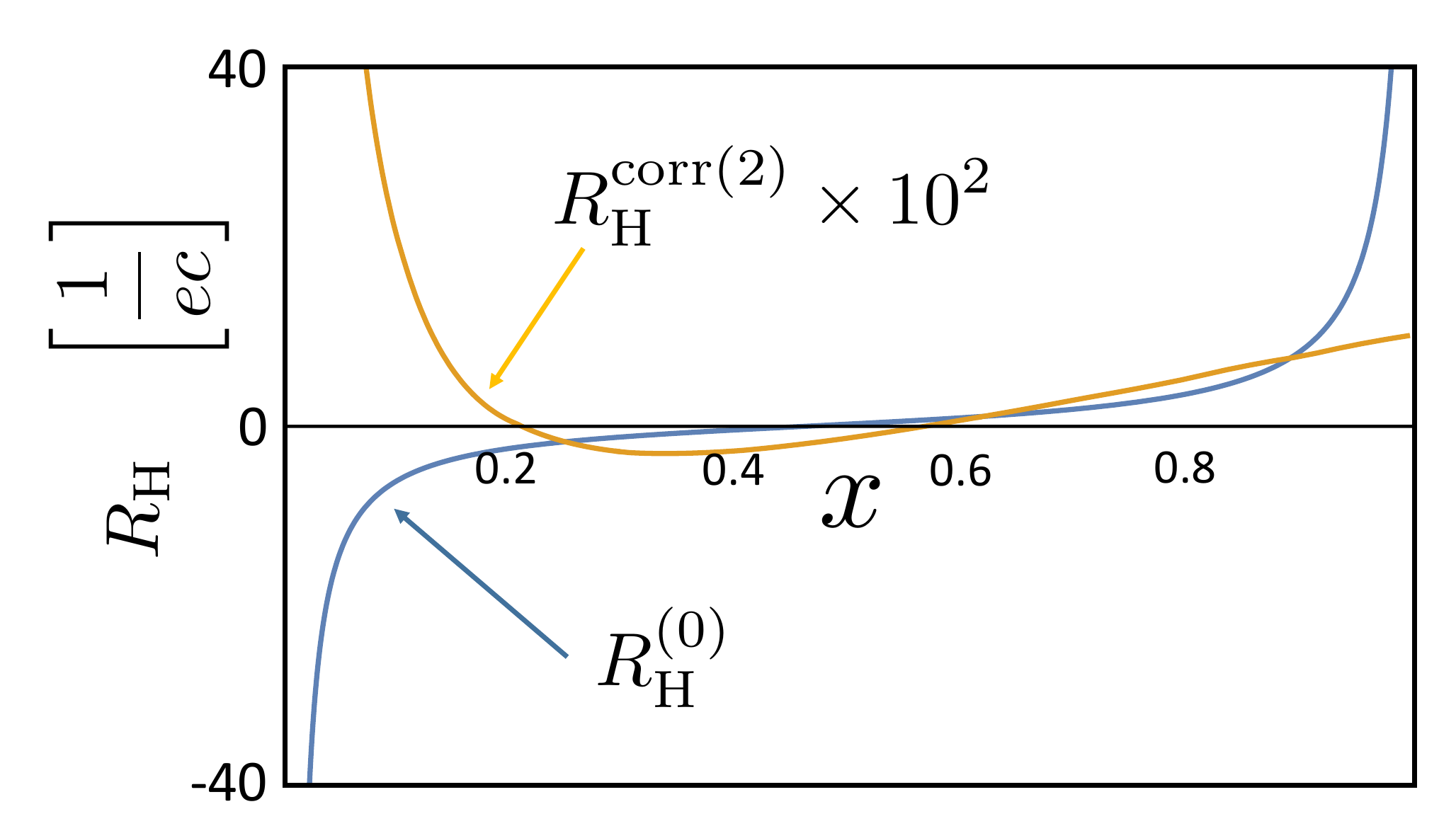}
\caption{Second order Hall coefficient correction compared to the zeroth order Hall coefficient, as defined in Eq.~(\ref{Chis}).   
The ratio of magnitudes  vanishes at $x\to 1$, and reaches $0.06$ at $x\to 0$. 
}
 \label{fig:RHcorr}
\end{center}
\end{figure}

Thus, the zeroth order Hall coefficient in the IT regime is provided analytically as a function of doping and temperature:
 \bea
&&R^{(0)}_{\rm H}=  {1\over ec} \left(  {-5  +10 x + 3 x^2\over  8 x(1-x)}  \right.+\nonumber\\
&&~~~ \left. (\beta \tt)^2   { -45 - 53 x + 145 x^2 + 225 x^3\over 192 e x}\right).  
\label{RH-t}
\eea
The $\beta\to 0$ limit of Eq.~(\ref{RH-t})  is depicted by the blue line in Fig.~\ref{fig:RHcorr}.

\subsection{Estimation of $R^{\rm corr}_{\rm H}$ at high temperature}
\label{sec:Corr}
In  Appendix \ref{App:Corr}, the correction term is fully defined as a sum which contains conductivity recurrents and hypermagnetization matrix elements between higher order Krylov operators created by nested commutators  $[[[H,H,\ldots ,j^\alpha]]]$.   
We calculate the contribution of  the first three terms,  
\be
R_{\rm H}^{{\rm corr}(2)} =  {1\over \chi_{\rm csr}} \left(  \left({\Delta_1\over \Delta_2}\right)^2 M_{2,2}'' -\left({\Delta_1\over \Delta_2}\right)\left( M_{2,0}''+M_{0,2}''\right) \right)
 \label{RHcorr2}
 \ee
 where  $\Delta_1,\Delta_2$ are the first two recurrents of the longitudinal conductivity, which are evaluated analytically in Appendix \ref{App:highTexp},
\be
\Delta_1^2=   t^2( 3  - 2x -x^2)  , 
~~\Delta_2^2 =  t^2 {24(1+x)\over 3 +x }   .
\label{recurrents}
\ee

In Fig.~\ref{fig:RHcorr}, $R_{\rm H}^{{\rm corr}(2)}$ is plotted as a function of doping, and compared to the zeroth order term $R_{\rm H}^{(0)}$.
The relative magnitude is qualitatively negligible  and 
is maximized toward $x\to 0$ where
 \be
\lim_{x\to 0}  \left| R_{\rm H}^{{\rm corr}(2) }   /R^{(0)}_{\rm H} \right| \to 6\% .
\label{ratio}
 \ee
Higher order correction terms $i,j \ge 2 $ in  Eq.~\ref{formula} consist of products of ratios of consecutive recurrents $\Delta_{2j-1} \over \Delta_{2j}$ times the hypermagnetization matrix elements $M''_{2i,2j}$. 
These terms are expected to be strongly suppressed relative to the low order terms due to the following argument: 
While generically the ratios of $\Delta_n/\Delta_{n+1}$  do not asymptotically decay rapidly with $n$~\cite{viswanath_recursion_book},   
$M''_{n,m}$  at high temperature are expected to diminish rapidly with $n,m$, since they involve traces over two  clusters of operators which are created by nested commutators  $\cL^m j^x=[H,[H,\ldots,j^x]]$  and $\cL^n j^y=[H,[H,\ldots,j^y]]$. 
The number of clusters in each of the normalized Krylov states increases faster than exponentially.
Since the clusters created from the $x$ and $y$ currents  occupy partially overlapping areas on the lattice,  the  fraction of operators which precisely match  the sites of $\tilde{c}_i,\tilde{c}_i^\dagger,s^\alpha_i$ decreases rapidly with the order of the normalized Krylov states.
This effect is proven already by the relative small size of  $R_{\rm H}^{{\rm corr}(2)} /R_{\rm H}^{(0)}$, and we expect the relative contributions of  the rest of the corrections to be  even smaller. 
 \begin{figure}[h!]
\begin{center}
\includegraphics[width=8cm,angle=0]{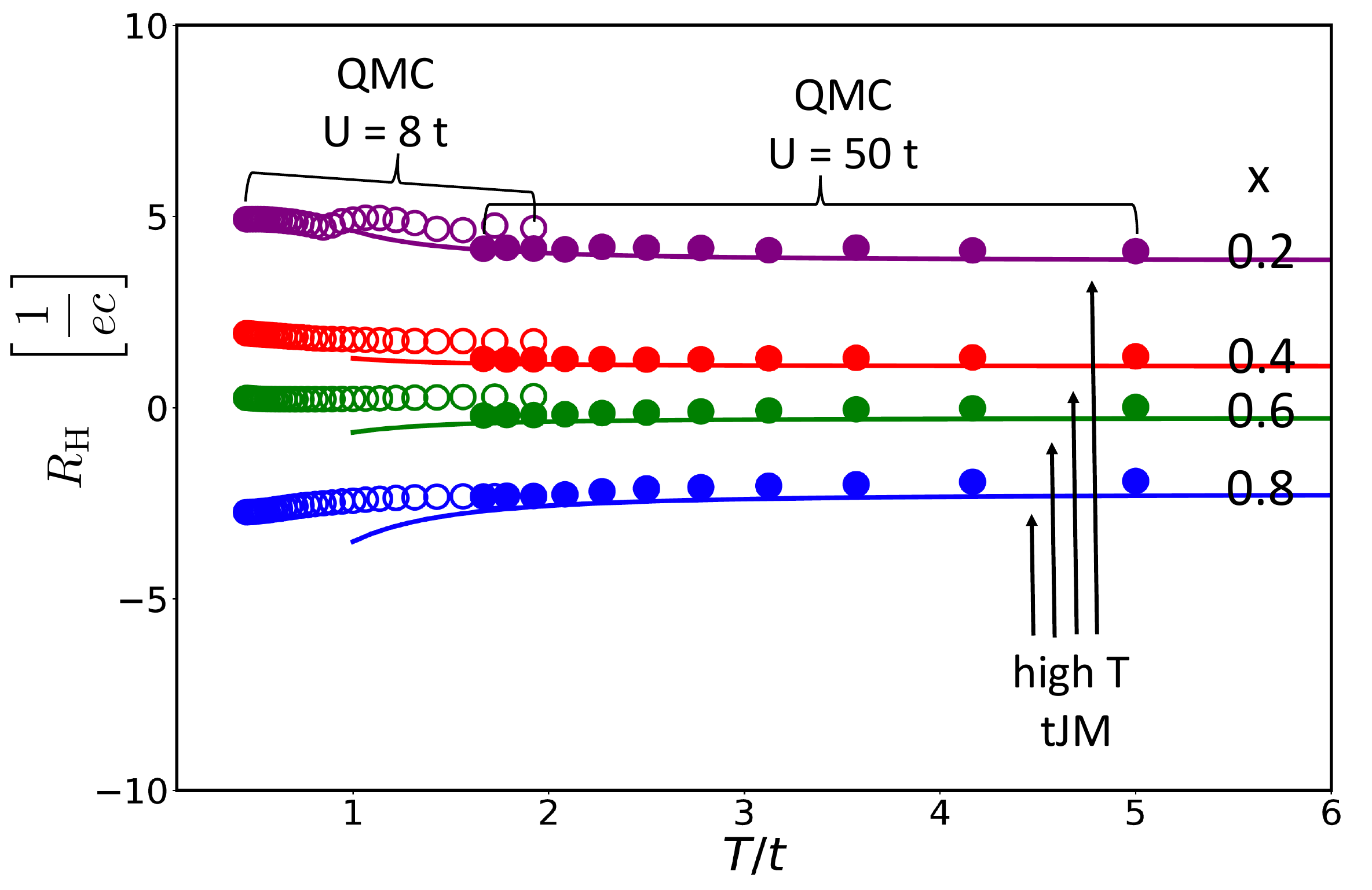}
\caption{
Hall coefficient $R_{\rm H}^{(0)}$  in the intermediate temperature regime $J < T < U$. Lines depict the high temperature expansion results, Eq.~(\ref{RH-t}). Solid and open  circles are QMC results using HM weights, 
for two values of $U/t$. The QMC results are plotted in the regions of negligible fermion sign error (see Appendix \ref{App:QMC-sign}). We note that the high temperature expansion agrees with the QMC data down to $T\simeq 2t$, and  
that $R_{\rm H}$ shows quite weak temperature dependence down to $T\simeq 0.5 t$.  } \label{fig:QMC-RH}
\end{center}
\end{figure}

\subsection{QMC extension to lower temperatures}
\label{App:QMC-sign}
 The QMC  extends the calculation of  $R^{(0)}_{\rm H}$ of the tJM
 to larger values of $\beta$. 
 
A  determinantal QMC calculation  for lattice fermions with discrete auxiliary fields was implemented using the ALF package~\cite{ALFQMC}.
We used  HM weights  for  
 $U/t=8,16,50$.
The typical system size was chosen to be $8\!\times\!8$,
with little size dependence, which was tested up to   size  $12\!\times\!12$, indicating
short correlation length in the studied temperature regime.
The imaginary time step was  chosen to render
 the Trotter errors to be  insignificant. 
The  number of Monte Carlo sweeps was generally $\sim 10^{5}$.
The statistics was quite well-behaved, and  {\em``Jackknife
resampling''}  (a method used for error estimation), revealed sufficiently small error bars. 
The  average sign in the QMC sampling is defined as 
\be
\langle S\rangle=\langle \sgn({\rm det})\rangle.
\ee
 In  Appendix \ref{App:QMCsign}, we report the value of $\langle S\rangle$ as a function of interaction strength $U/t$, doping 
and temperature. We show that quite generally,  $\langle S\rangle$ approaches unity at higher temperatures where the Fermionic negative weights introduce negligible effects on QMC configuration averaging.
 
The  CMC and CSR susceptibilities of Eq.~(\ref{Chis}) were computed  
by sampling products of Green's functions using Wick's theorem 
over QMC equilibrium configurations of the auxiliary fields. 
In  Fig.~\ref{fig:QMC-RH} the QMC  results are depicted by circles of larger diameter than the numerical error bars.  
The displayed data is restricted to the regime of $\langle S\rangle \ge 0.8$, which for  $U=8t$ and all doping range is satisfied at $T \ge  t/2\approx J$.
We note that the data  exhibits a weaker temperature dependence 
than expected by extrapolating the analytic high temperature results.

\section{$R_{\rm H}$  at very high temperatures}  
\label{sec:VHT}
At very high temperatures  $T> U$,
$R_{\rm H}$ for HM  is obtained by a high temperature expansion  in powers of $\beta U\ll 1$.  
The commutators between  unprojected magnetization and currents  do not involve interaction terms, and are bilinear in fermion operators.  The  leading orders in the high temperature expansion are given by traces over these operators,
\bea
&&  \chi^U_{\rm csr}\sim \beta e^2 t^2 n(2-n)   , \quad
 \chi^U_{\rm cmc}\sim   \beta^2 {e^3\over c} t^4 n(2-n)(1-n)   \nonumber\\
&& R^{U}_{\rm H} =  {2(1-n) \over   n(2-n)ec} +\cO(\beta U)^2 ,
\label{RH-HM}\eea
where $n=1-x$ is the electron density.  Thus, the leading order  in
$R^U_{\rm H}$ recovers  the high temperature expansion of the non-interacting  square lattice  coefficient which is depicted in Fig.~\ref{fig:HallMap-SL}.

Interestingly, this effect is qualitatively implemented by the addition of the  next neighbor hopping term $H^{J} $ of order $J$  in the tJM. 
As a hopping term, $H^{J}$ in Eq.~(\ref{tJM})  contributes terms of order  $J\ll t$ to the  current  and magnetization operators,
\bea
&&{j'}^{\alpha}_{ijk}= -i eJ (1-n^h_j) \left( K^-_{ik;\alpha}-   2{\bf \Sigma}^-_{ik;\alpha}\cdot \bs_j\right)   \nonumber\\
&&M' =   {1\over 2c}   \sum_{\ijk} ( x_i {j'}^{y}_{  ijk}- y_i  {j'}^x_{ ijk}  ).
\label{OrderJ}
\eea
Since $H^{J}$ connects sites across the plaquette diagonals, its high temperature expansion   yields a power of $\beta J$, which is  one power lower than the leading power of $ \chi^t_{\rm cmc}$,
(see Appendix~\ref{App:HJ}). Thus,  
\be
\chi^{tJ}_{\rm cmc} = \chi^t_{\rm cmc}  +  {\beta J e^3\over 2} x(1-x)(1+2x-3x^2) (1+ \cO(t/U) ).
\ee
Combining the Hall coefficients of Eq.~(\ref{RH-t}) with the contribution of  $H^{J}$ to the second order in $\beta \tt$, and neglecting the correction term and terms of order $t/U$ yields
\be
R^{tJ}_{\rm H}= R^{t}_{\rm H} + {1\over ec}  {4 \over \beta U} {(1+2x-3x^2)  \over 8  ex (1-x)} .
\label{RH-tJ}
\ee

The additional   term  is opposite in  sign to $R_{\rm H}^t$. For $\beta U  <  4$  (beyond the validity of the tJM),  $R_{\rm H}$  is expected to become negative
as depicted in Fig.~\ref{fig:HallMap-U}.
 
\section{Resistivity slope}
  \begin{figure}[h]
\begin{center}
\includegraphics[width=8cm,angle=0]{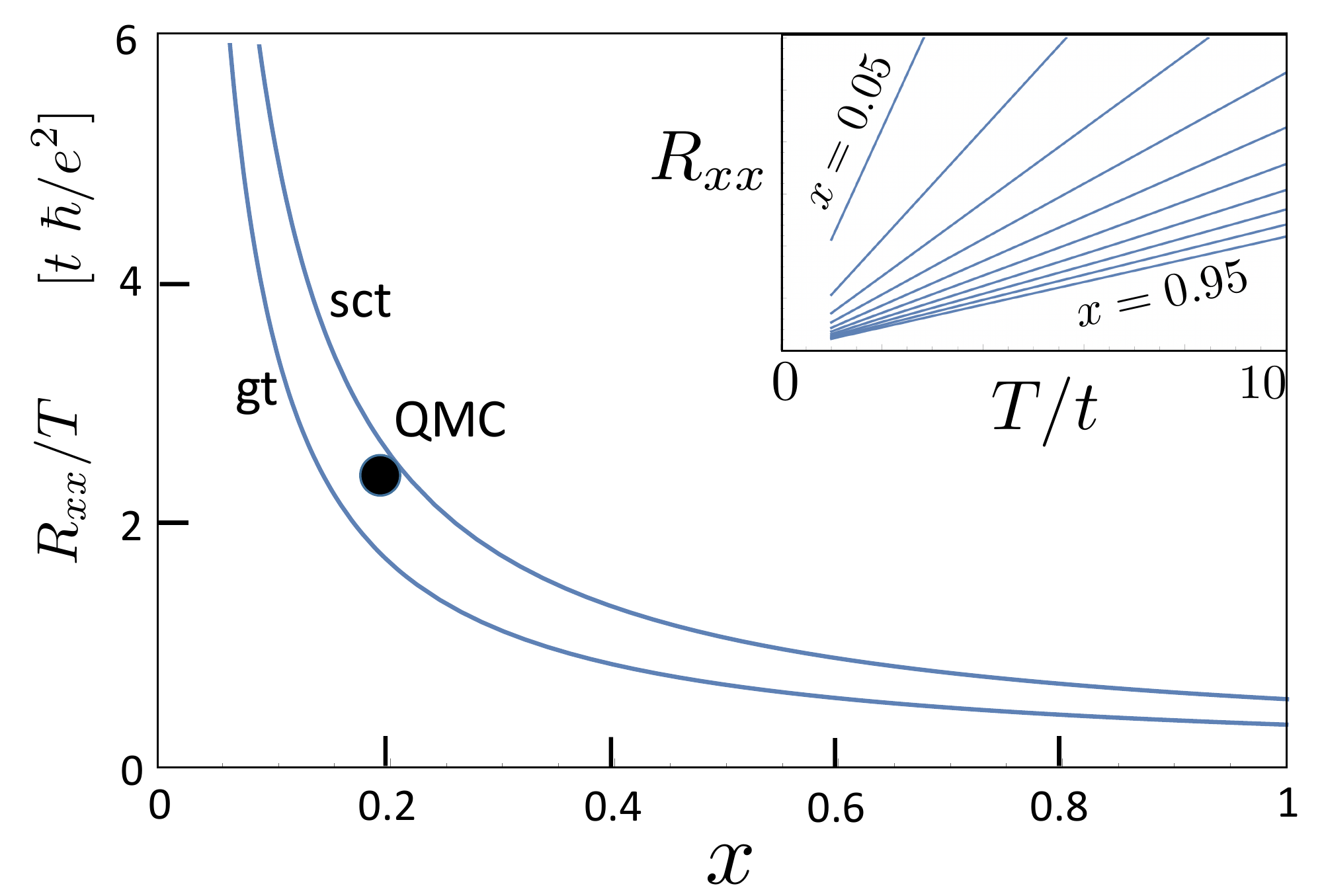}
\caption{High temperature resistivity slopes of  $H^t$ as a function of doping  $x$. 
SCT and  GT denote two different continued fraction extrapolation schemes  (see text).  Inset depicts  $R_{xx}(T)$ for the GT.
The solid circle marks the QMC result for the $U\!=\!8t$ HM, reported in Fig. S3 of Ref.~\cite{Dev-Rxx}.  \label{fig:RXXvsT} }
\end{center}
\end{figure}

The continued fraction expression for the longitudinal  conductivity~\cite{LA,Ilia} is given by
 \be
 \sigma_{xx}(\omega) =\beta \chi_{\rm csr} \Im {1 \over    -\Delta_1^2 G_2( -i\varepsilon)}
 \ee
 where $\Delta_1,\Delta_2$ and $G_2$ are  the first two conductivity recurrents, and the estimated second order termination function respectively.
$G_2$ is estimated by two  extrapolation schemes. First, we use the semicircle termination (SCT) where all higher order recurrents are assumed to be equal to $\Delta_2$, 
  \be
 \Delta^{\rm sct}_n=\Delta_2, \quad n=2,3\ldots \infty. \ee
This yields an algebraic equation for $G_2$, 
 \bea
 G^{\rm sct}_2( -i\varepsilon) &=& {1\over  -i\varepsilon  -\Delta_2^2 G_2( -i\varepsilon)},\nonumber\\
\Im G^{\rm sct}_2 &=&   -{1\over \Delta_2}, \nonumber\\
 \sigma^{\rm sct}_{xx}&=& \beta \chi_{\rm csr}  {\Delta_2\over\Delta_1^2},\nonumber\\
R^{\rm sct}_{xx}&=& T {\Delta_1^2 \over  \chi_{\rm csr}  \Delta_2 }=   {T\over t}{(3+x)^{3\over2} \over 4\sqrt{6} x \sqrt{1+x} }.
\label{SCT}
 \eea
 
Second, we use the Gaussian termination (GT), which assumes that the recurrents $\Delta_{n\ge 2}$ scale as $\sqrt{n}$,
\be
 (\Delta^{\rm gt}_n)^2 ={1\over 2} n \Delta^2_2, \quad n=2,3,\ldots \infty. \ee
This extrapolation yields,
 \bea
\Im G^{\rm gt}_2(0)&=&   -{2\over \sqrt{\pi}  \Delta_2} ,\nonumber\\
 \sigma^{\rm gt}_{xx}&=& \beta  {\sqrt{\pi}\over 2} \chi_{\rm csr}   {\Delta_2\over\Delta_1^2}.
 \eea

We note that the two different extrapolations yield similar results for the DC resistivity $R_{xx}=\sigma_{xx}^{-1}$:
\be
R^{\rm gt}_{xx}=   {2\over \sqrt{\pi}} R^{\rm sct}_{xx}.
\label{GT}
\ee
In Fig. \ref{fig:RXXvsT}, the resistivity slopes of  $H^t$  are plotted using the SCT and GT extrapolations. We note that the slopes diverge toward the Mott limit, as expected by the suppression of the CSR depicted in Fig.~\ref{fig:CSR}.
Interestingly, the  resistivity is finite at high temperatures even in the dilute electron density limit   $x\to 1$. We note a quantitative agreement of the slope with the calculation of HM recurrents in Ref.~\cite{Dev-Rxx}.

 \section{Discussion}
A relevant precursor  to our work includes  the calculation
of the high frequency limit of the Hall coefficient of the tJM by  Shastry, Shraiman and Singh~\cite{SSS}.  It is   interesting (but far from obvious) that they have found 
a qualitatively similar  doping dependence to the zero frequency Hall coefficient calculated here. We also note  that sign reversals and Hall coefficient increase towards the Mott limit
have been obtained in some parameter ranges of the HM using dynamical mean field theory~\cite{HallDMFT,Dev-PR,Khomskii}.

The formula for $R_{\rm H}^{(0)}(x,T)$  given in Eq.~(\ref{Chis}) was computed by QMC for the HM~\cite{Dev-Rxy,Dev-PR}.  A Hall coefficient sign reversal  and increase toward the Mott phase was detected albeit with a much reduced magnitude
relative to the tJM calculation.  This difference is attributed to the use of the HM for large $U/t$ instead of the lower energy effective tJM.  The CSR of the HM includes dynamical  conductivity contributions above the Hubbard gap. Hence, the denominator of $R_{\rm H}$ does not vanish at zero doping, and it does for the tJM.  $R_{\rm H}$ can be calculated in principle using either HM or tJM. However, since $R_{\rm H}^{\rm corr}$ depends on $[H,j^x]$, the HM whose current-Hamiltonian commutator scales with the interaction strength $U$,
produces  a larger correction than  the tJM.

Our results lead to the following conclusions:

(i) Strong interactions which open a Mott gap at zero doping, also affect the sign and density of charge carriers in a sizeable portion of the  Hall map  as depicted in Fig.~(\ref{fig:HallMap-U}).

(ii)  $R_{\rm H}$ and  $R_{xx}/T$ increase as $x\to 0$ due to the suppression of the CSR toward the Mott insulator.  

(iii) The spin-charge correlated commutators of the GP currents of  Eq.~(\ref{comm}), are the source of the Hall sign reversal at finite doping.

At lower temperatures than discussed in this work, one still expects the strong interactions to have implications on the charge transport.
If, as numerically predicted~\cite{DMRG,Troyer,Sorella},  $d$-wave superconductivity emerges at low doping of the tJM, its condensate should be described by a low density of GP hole pairs, rather than the Cooper pairs on a narrow shell on the putative
band-theory predicted Fermi surface. 

The experimental manifestation of the positive constituent charges in cuprates is found in  Hall conductivity above and below the superconducting temperature.
Bardeen and Stephen theory predicts the same Hall sign in the flux flow regime as in the normal phase~\cite{BS}. At lower temperatures, the Hall conductivity acquires an additional negative contribution from the vortex charge~\cite{SciPost}, which depends on the
derivative of the superfluid stiffness $\rho_s$ with respect to electron density $n_e$.  The negative sign of $d\rho_s/dn$ at low doping~\cite{Uemura} is consistent with  a condensate of positively charged hole pairs.

{\em  Acknowledgements} -- We thank Omer Yair who worked on a predecessor of this project, and Netanel Lindner, Edward Perepelitsky, Sriram Shastry, and Efrat Shimshoni for useful discussions. 
A.S. and S.B thank Fakher Assaad and Johannes Hoffmann for their help in application of the ALF numerical packages. We acknowledge the
Israel Science Foundation Grant No. 2021367.
This work was performed in part at the Aspen Center for Physics,  supported by National Science Foundation grant PHY-1607611,
and at Kavli Institute for Theoretical Physics, supported  by Grant No. NSF PHY-1748958.

\bibliographystyle{unsrt}
\bibliography{refs.bib}

\begin{thebibliography}{10}
\bibitem{Comm-FLT}
{One assumes that there are no translationally broken symmetries at the
  measured temperatures, which could change the volume of the Fermi surface}.
\bibitem{Comm:QMC-AC}
{Imaginary time QMC conductivities require analytic continuation, which is
  limited to frequencies higher than temperature (see Appendix B in
  \cite{Snirgaz}). The DC conductivities are often deduced by proxies to the
  analytical continuation~\cite{Dev-PR}.}
\end{thebibliography}

%\newpage
\appendix
\section{High temperature expansion of the susceptibilities}
\label{App:highTexp}
In the grand cannonical ensemble,  the mean density of holes for the tJM can be imposed at infinite temperature by a product density matrix $\rho_0(\delta)$, with a fugacity parameter $\delta$:
\be 
\rho_0(\delta)= \prod_i   \left(  \delta  |h_i \rangle\langle h_i |+\sum_{s=\uparrow ,\downarrow} {1-\delta \over 2} |s_i\rangle\langle s_i|   \right),
\label{Tr1}
\ee 
where $|h_i\rangle$ is a hole at size $i$, and the average hole density at  infinite temperature is
\be
x(0)=\langle n^h_i\rangle_{\beta=0} \equiv  \delta .
\label{Tr2}
\ee
Expectation values are,
\bea
\Tr \rho_0 s_i^\alpha (1\!-\!n^h_i)&=&0,\nonumber\\
\Tr \rho_0 \bs^2_i  (1\!-\!n^h_i) &=&{3\over 4}(1-\delta) .
\label{Tr3}
\eea
Using the relation,
\be
K^{s,a}_{12} K^{s,a}_{21}=n^h_1+n^h_2 - 2 n^h_1 n^h_2,
\label{K2}
\ee
we obtain some useful traces on two-site operators:
\bea
\Tr \rho_0 (K_{12}^+)^2 &=& - \Tr (K_{12}^-)^2 =2\delta(1-\delta)\nonumber\\
\Tr \rho_0 \left( \Sigma^+_{12,\alpha} \right)^2  &=&-\Tr \left( \Sigma^-_{12,\alpha} \right)^2 =2\delta(1-\delta)\nonumber\\
\Tr \rho_0 s^\alpha_i s^\beta_j &=& \delta_{ij}\delta_{\alpha\beta} {1\over 4}.
\label{K3}
\eea

At finite temperatures,  we  expand the average hole density $x(\delta)$  to second order in $\beta$,
\bea
x    &=& {\Tr \rho_0(\delta) e^{-\beta H}n^h_i  \over \Tr   \rho_0(\delta_0) e^{-\beta H} } = \delta + {\beta^2\over 2} \Tr \rho_0(\delta) H^2 (n^h_i - \delta )\nonumber\\
&=& \delta  +  2  \beta^2 (1- 2 \delta) \delta (1-\delta ) +\cO(\beta^4).
\label{dens-corr}
\eea
Eq.~(\ref{dens-corr})  allows us to evaluate the  density dependence of  a $\delta$-dependent susceptibility by
\be
\chi(\delta, \beta) = \chi (x-2  \beta^2 (1- 2 \delta) \delta (1-\delta ),\beta).
\label{corr}
\ee
 
 \subsection{CSR}
 \begin{figure}[h]
\begin{center}
\includegraphics[width=7cm,angle=0]{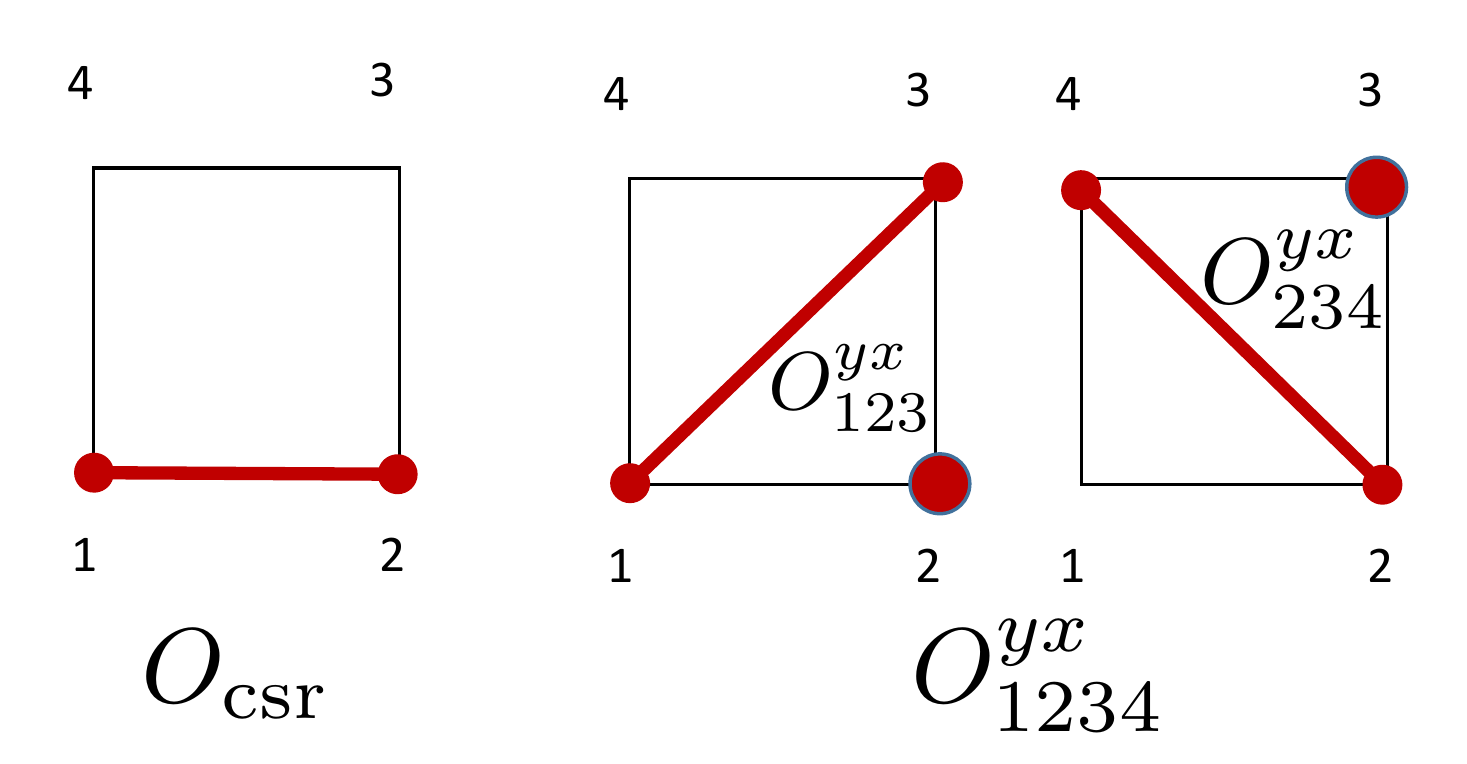}
\caption{Operators which are averaged over in the CSR and CMC. Line denotes a bond operators $K^+$ or ${\bf \Sigma}_\alpha^{+s}$, 
and the circle  denotes a site operator ${1+n^h\over 2}$ or  $s^\alpha (1-n^h)$, respectively.  }
 \label{fig:Operators}
\end{center}
\end{figure}

  The first two leading orders in $(\beta t)$ of the CSR are defined as,
 \be
\bar{ \chi}_{\rm csr} = e^2 (\beta \tt) \bar{\chi}^{(1)}_{\rm csr} + e^2(\beta\tt)^3 \bar{\chi}^{(3)}_{\rm csr} +\ldots.
 \ee
where, using (\ref{K2}), 
\be
\bar{\chi}_{\rm csr}^{(1)} =    \Tr \rho_0 ( K^+s_{12})^2 =   2x(1-x) .
 \ee
The  order $\beta^3$ CSR is obtained by expanding the Boltzmann weight and the partition function,
 \be
\bar{ \chi}_{\rm csr}^{(3)}(\delta)  =   {1\over 6}  \Big( \overbrace{ -\Tr_c \rho_0 ((H/t)^3 K^+s_{12}   ) }^{A} - 21 \overbrace{  \left(\Tr  \rho_0 (K_{12}^+s)^2 \right)^2  }^{B} \Big),
\label{CSR-3}
\ee
\begin{figure}[h]
\begin{center}
 \includegraphics[width=8cm,angle=0]{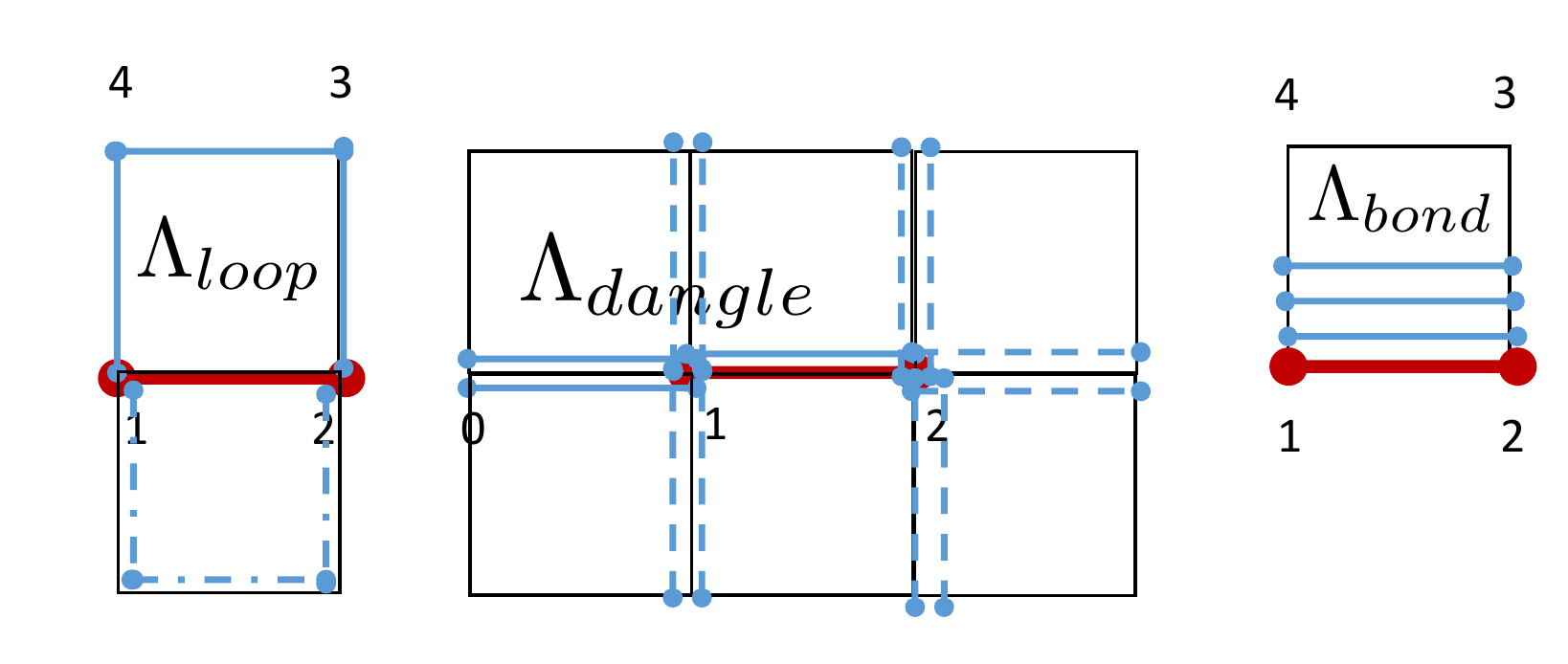}
\caption{Graphs  of operators which contribute to the term $A$ in Eq.~(\ref{CSR-3}). }
 \label{fig:CSR-3}
\end{center}
\end{figure}

where $c({12})$ includes all the $H$-bonds emanating from sites 1,2, as depicted in Fig.\ref{fig:CSR-3}. The traces yield,
\bea
    &&A=   \delta(1-\delta) \left( 15 -10\delta + 13\delta^2\right),\nonumber\\
&&  B =        84 \delta^2(1-\delta)^2,\nonumber\\
&& \bar{ \chi}_{\rm csr}^{(3)}(\delta) =    {1\over 6} \delta(1-\delta)   \left(15- 94\delta + 97\delta^2 \right).
\eea
 Using Eq.~(\ref{corr}) to transform $\chi_{\rm csr}(\delta)$ to $\chi_{\rm csr}(x)$ yields, 
 \bea
  \chi_{\rm csr}^{(3)} (x)  &=&     \bar{\chi}_{\rm csr}^{(3)} (x ) - {\partial  \bar{\chi}_{\rm csr}^{(1)} \over \partial x } 2  (1- 2 x) x(1-x)  \nonumber\\
  &=& {1\over 6}  x(1-x) (-9 + 2 x + x^2 ).
\eea
This recovers Eq.~(\ref{CSR-hiT}), which was previously published in Ref.\cite{Perepel}. 
  
 The  CMC is given by averaging the plaquette operator, shown in Fig.~\ref{fig:Operators}.
\bea
\chi_{\rm cmc} &=&    -  2 e^3 \tt^3 \langle O^{yx}_{123}\rangle \nonumber\\
&=&    2  e^3 \tt^3\langle  \left(  K^+s_{13} { 1+n^h_2 \over 2} + {\bf \Sigma}^+s_{13}  s^\alpha_2 (1-n^h_2) \right)\rangle ,\nonumber\\
\eea
where we have used using C4 symmetry to equate  four identical contributions to the  expectation value. 
The leading order $\chi^{(2)}_{\rm cmc}$ requires tracing $ O^{yx}_{123}$ times two Hamiltonian bonds in a connected cluster inside a plaquette.
The calculation yields 
\bea
 \chi^{(2)}_{\rm cmc} &=&  -  \Tr \rho_0  (\beta H)^2 O^{yx}_{123} \nonumber\\
&=&  {\beta^2 \over 2 }   x(1-x)  (-5+10x  + 3x^2 ).
 \label{chiCMC-2} 
 \eea

\begin{figure}[h]
\begin{center}
\includegraphics[width=6cm,angle=0]{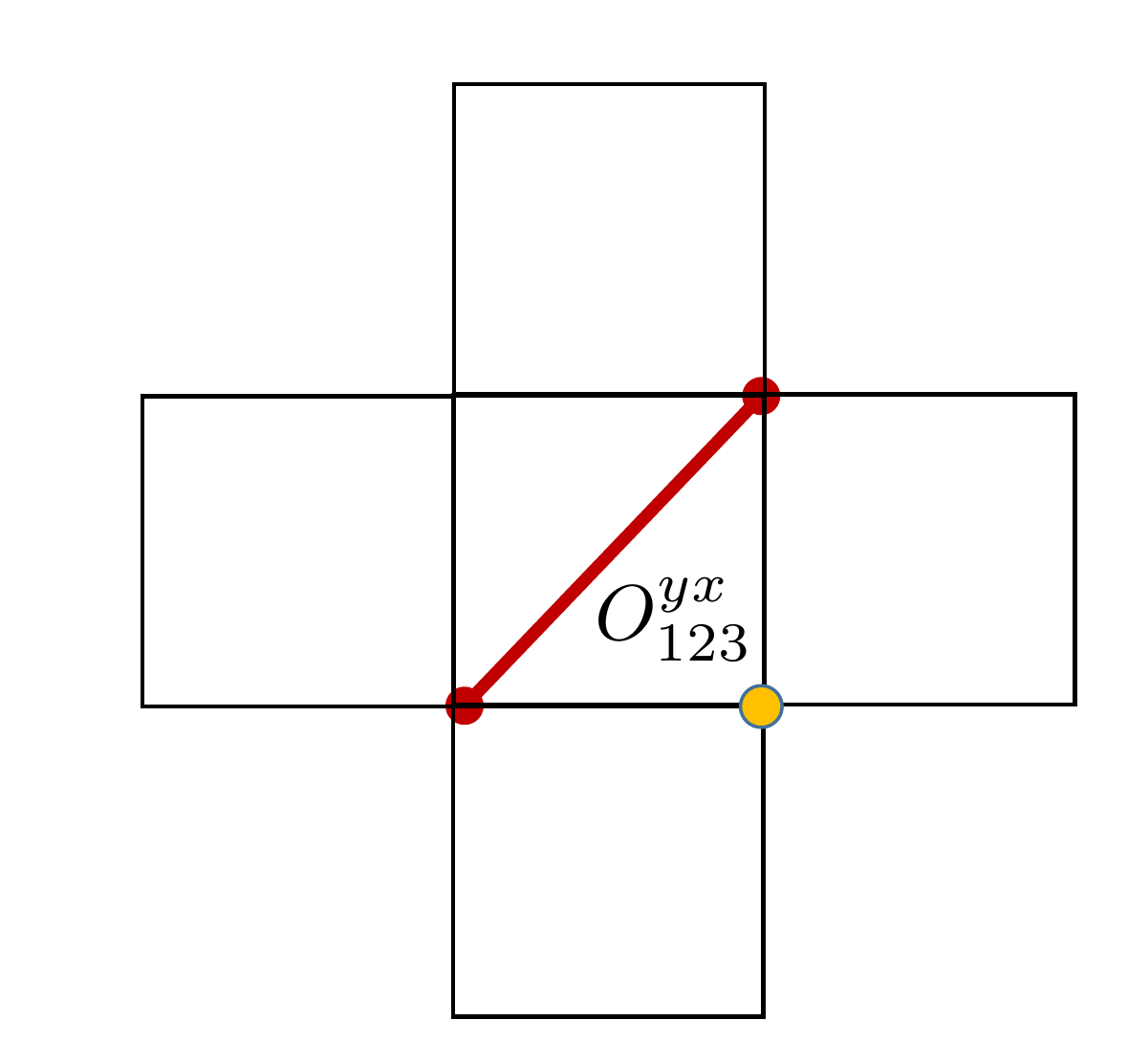}
\caption{Minimal connected cluster for the   calculation of   $\chi^{(4)}_{\rm cmc}$, which must include 4 powers of $H$ whose bonds connect to the sites of $O^{yx}_{123}$
and contribute to  a non vanishing trace.}
 \label{fig:CMC-4}
\end{center}
\end{figure}
The order $\beta^4$  is given by
\bea
&&\chi_{\rm cmc}^{(4)} (\delta) =   -{1\over 12}  \Tr_c \rho_0 (\beta H/t)^4 O^{yx}_{123} )  \nonumber\\
&&~~~~~~ - 8  \delta^2 (1-\delta)^2     (-5+10\delta  + 3\delta^2 ) .
\label{chiCMC-4}
\eea
The first term  in (\ref{chiCMC-4}) which is calculated on the lattice shown in Fig.~\ref{fig:CMC-4}, is computed   by symbolic multiplication.  
\bea
&&\Tr_c \rho_0 ( (H/t)^4 O^{yx}_{123} )  ={1\over 4}  \delta (1-\delta ) \nonumber\\
&&  ~~~~~~~~ \times \left(213\delta^4 + 1104\delta^3 -2022\delta^2 + 408\delta + 105 \right) .\nonumber\\
\eea
This results in,
\be 
\chi_{\rm cmc}^{(4)} (\delta) ={1\over 16}  \delta(1-\delta) (-35 + 504 \delta - 1246 \delta^2 + 528 \delta^3 + 313 \delta^4).
\ee
Transforming the expression using Eq.~(\ref{corr}), yields
\bea
\chi_{\rm cmc}^{(4)} (x)&=& \lim_{\delta\to x}\{  \chi_{\rm cmc}^{(4)} (\delta) -(\partial_\delta \chi^{(2)}_{\rm cmc})  2 (1 - 2 \delta) \delta (1 - \delta)\} \nonumber\\
&=& {x(1-x)\over 16}  (45 - 136 x + 50 x^2 +48 x^3 - 71 x^4),\nonumber\\
\eea
which arrives at  Eq.~(\ref{CMC-hiT}). 

\section{Leading order effect of $H^{J}$} 
\label{App:HJ}
The correlated hopping term in $H^J$ of Eq.~(\ref{tJM})  closes a loop of three bonds on the square lattice, which produces terms of order $(\beta J) $ in the CMC.
Collecting  the leading orders in $(\beta J)$ 
from $H^{J}$, the currents and magnetization yields
\be 
\chi_{\rm cmc}^J  =   2 e^3 \beta J t^2   \beta (C_1 + C_2),
\ee
where
\be 
C_1  \equiv   \Tr \rho (H^{J}/J) O^{yx}_{123}   =   - {1 \over 4  }   \delta (1-\delta)^3  ,
\label{C1}
\ee 
and
\bea C_2 &\equiv &   \Tr \rho_0 K_{14}^+  \left[  - n^h_4 , [ j^J_{234} /J, j^x_{12}] \right]    \nonumber\\
&&  +\Tr \rho_0   K_{14}^+   \left[  - n^h_4 , [ -j^x_{43} , j^J_{123}/J] \right] ,\nonumber\\
&=&     2 \times {1 \over t} \Tr \rho_0  (K_{14}^+s)^2 {1+n^h_2\over 2}    (1-n^h_3) \nonumber\\
&=  & {1 \over 2 }  x (1- x )(1- x ^2) .
\eea
Thus,
\be
\chi_{\rm cmc} =  \chi^t_{\rm cmc}+ {\beta J e^3 t^3 \over 2} x(1-x)(1+2x-3x^2) ,
\ee
which yields expression (\ref{RH-tJ}). 
The order $J=4t^2/U$ becomes  as large as $\chi_{\rm cmc}^t$ at around $T\ge U$, which is where the renormalization of the HM onto the tJM ceases to be valid.

 \section{Second order corrections of the Hall coefficient formula}
 \label{App:Corr}
 At high temperatures, the scalar product  between operators is given by
 \be
 (A|B) = \beta \Tr A^\dagger B + {\beta^2\over 2} \Tr \{H,A^\dagger \} B +\cO(\beta^3).
 \label{eq:scalarprod}
 \ee
   
 The Hall coefficient correction~\cite{EMT},
 \bea
R_{\rm H}^{\rm corr}&=& {1\over \chi_{\rm csr}}  \sum_{i,j=0}^\infty R_i R_j  (1 -\delta_{i,0} \delta_{j,0})M''_{2i,2j}\nonumber\\
M''_{2i,2j}&=&  \Im \left(  \langle (2i)_y |  \cM |(2j)_x\rangle] - \langle (2i)_x | \cM |(2 j)_y\rangle] \right)  ,\nonumber\\
R_{i>1} &=& \prod_{r=1}^i \left(-{\Delta_{2r-1}\over \Delta_{2r} } \right)\quad R_0 =1.  
\label{formula}
\eea

The recurrents are defined as $\Delta_0=0$,
\be
\Delta_n \equiv  \langle  (n)_x |\cL (n+1)_x\rangle, n \in \mathbb{Z},
\ee
which can be obtained from the lower order conductivity moments $\mu_{2k}$ of orders $k\in \mathbb{Z}$, where
\be
\mu_{2k} = (j^x | \cL^{2k} |j^x) = - \Im \langle [j^x,\cL^{2k-1} j^x]\rangle,
\ee
where the Liouvillian hyperoperator is defined as $\cL O=[H,O]$, $j^\alpha = i\left[ H,P^\alpha \right]$ and  $\mu_0=\chi_{\rm csr}$.

The  magnetization hyperoperator is $\cM O=[M,O]$.  The  orthonormal Krylov states  $|(n)_\alpha\rangle$, which  define the matrix elements $M_{2i,2j}''$ are a set of operators (hyperstates)  
created by applying $\cL^n$ to the current $\cL^n |j^\alpha\rangle = \left[H,\left[H,...\left[H, j^\alpha \right] \right]\right]$ \cite{EMT}, and  orthonormalizing  the result
with respect to the lower order  states. The Krylov basis is thus constructed as,
\be
|(0)_\alpha\rangle= { |j^\alpha)\over \sqrt{\chi_{\rm csr}}}, 
\ee
and
\be
|(n+1)_\alpha\rangle = {\cL |(n)_\alpha\rangle-\Delta_n |(n-1)_\alpha\rangle \over (\langle (n)_\alpha| \cL^2 |(n)_\alpha\rangle - \Delta_n^2  )^{1\over 2} },\quad n\ge 1.
\label{eq:krylov}
\ee

The second order correction terms of  Eq.~(\ref{RHcorr2}) include two recurrents $\Delta_1,\Delta_2$ and three  the hypermagnetization matrix elements 
$M_{0,2}'',M_{2,0}'',M_{2,2}''$.

\subsection{$\Delta_1$}
    \begin{figure}[h!]
\includegraphics[width=6cm,angle=0]{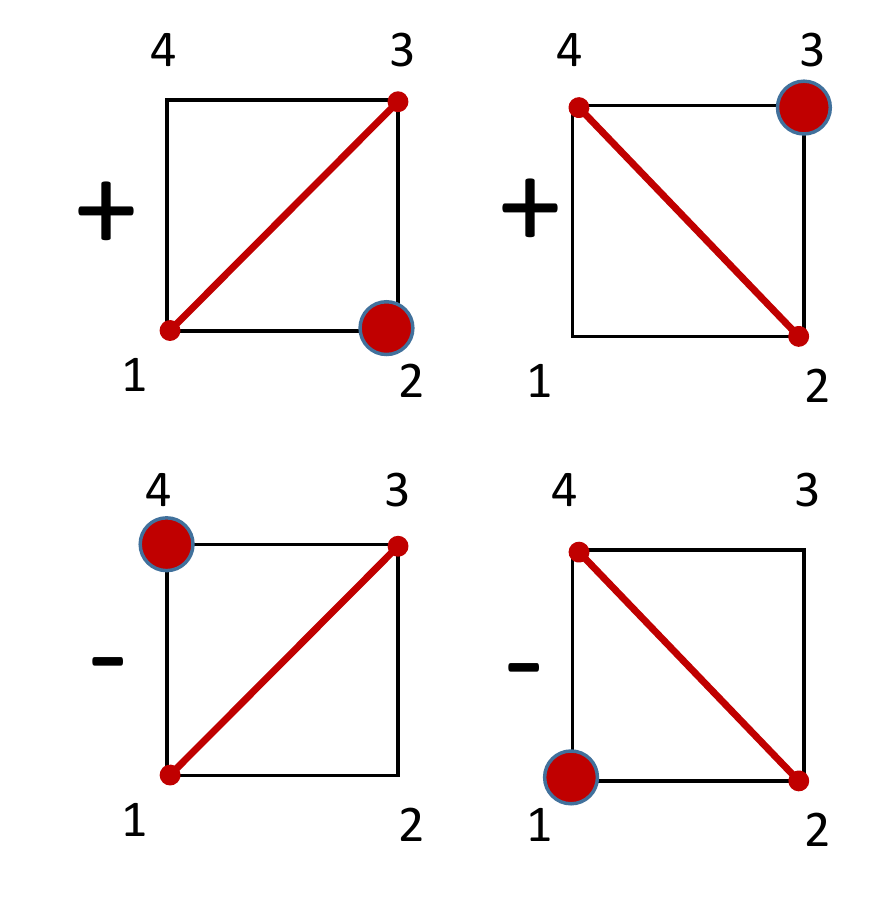}
 \caption{Operators of first unnormalized Krylov hyperstate $|(1)_x)=[H^t,j^x]$.  Solid line and circle represent the product of bond  and site  operators $K_{ij}^+s$, $(1+n_k^h)/2$ respectively, or
$\Sigma^+s_{\alpha,ij}$ and
 $\bs_k^\alpha(1-n_k^h)$.   \label{fig:Lj} 
}
\end{figure}
The first Krylov state $\cL j^x$ and $\cL j^y$ is described by the diagrams of Fig. \ref{fig:Lj}, 
\bea
\cL j^x &=& \sum_i C_{13}^x \left( K^+s_{13} {(n^h_2-n^h_4)\over 2} + \Sigma_{13,+s}^\alpha (s^\alpha_2-s^\alpha_4)\right)\nonumber\\
&&+C_{42}^x \left( K^+s_{42} {(n^h_3-n^h_1)\over 2} + \Sigma_{42,s}^\alpha (s^\alpha_3-s^\alpha_1)\right) \nonumber\\
\cL j^y &=& \sum_i C_{13}^y \left( K^+s_{13} {(n^h_2-n^h_4)\over 2} + \Sigma_{13,s}^\alpha (s^\alpha_2-s^\alpha_4)\right)\nonumber\\
&& +C_{42}^y \left( K^+s_{42} {(n^h_3-n^h_1)\over 2} + \Sigma_{42,s}^\alpha (s^\alpha_3-s^\alpha_1)\right) ,\nonumber\\
\label{Lj}
\eea
where the coefficients
\be
(C_{13}^x, C_{42}^x)=(1,1),\quad (C_{13}^y, C_{42}^y)=(-1,1),
\label{Csymm}
\ee
depend on the symmetries of $j^x,j^y$ respectively.
 
The first recurrent is defined by the norm of the operator $\cL j^x/\chi_{\rm csr}^{1\over 2} $:
  \bea
\Delta_1^2 &=&  { \beta\over \chi_{\rm csr}} \Tr j^x\cL^2 j^x\nonumber\\
&&= {   t^2 \over \chi_{\rm csr} }  2 \Tr \rho_x  \left( K_{13,a} K_{31,a} {(n^h_2 - n^h_{4})^2\over 4} \right. \nonumber\\
&& +\left.  2 \times 3  \Sigma^z_{13,a} \Sigma^z_{31,a}   (s^z_2)^2 \right) \nonumber\\
 &=&  { \beta\over  \chi_{\rm csr} } t^2 2x (1-x ) \left( x (1-x ) + 3(1-x )\right) =   3  - 2x  -x ^2 ,\nonumber\\
\eea
which is given in  the main text, Eq.~(\ref{recurrents}). Note that $\Delta_1^2$ is positive at all dopings, and vanishes at $x \to 1$ due to absence of scattering in the empty band.
Notice that near the Mott phase,  $x \to 0$,  $\Delta_1$ is dominated by the $({\bf \Sigma}\cdot {\bf s})^2$ term, representing scattering of holes from spins.

 \subsection{$\Delta_2$}
 The second recurrent is given by the equation, 
 \be
 \Delta_2^2 = { \mu_4\over \Delta_1^2 \chi_{\rm csr} } - \Delta_1^2 =   { \mu'_4\over \Delta_1^2 \chi_{\rm csr} },
 \ee
 where the fourth moment is
 \be
 \mu_4= \beta \Tr \rho_x   j^x   \cL^4 j^x ,
 \ee
 which contains the traces of the squares all non-returning ($\cL^2 j^x\propto j^x$) ) operators in $\cL^2 j^x$.  
 
\begin{figure}[h]
\begin{center}
\includegraphics[width=8cm,angle=0]{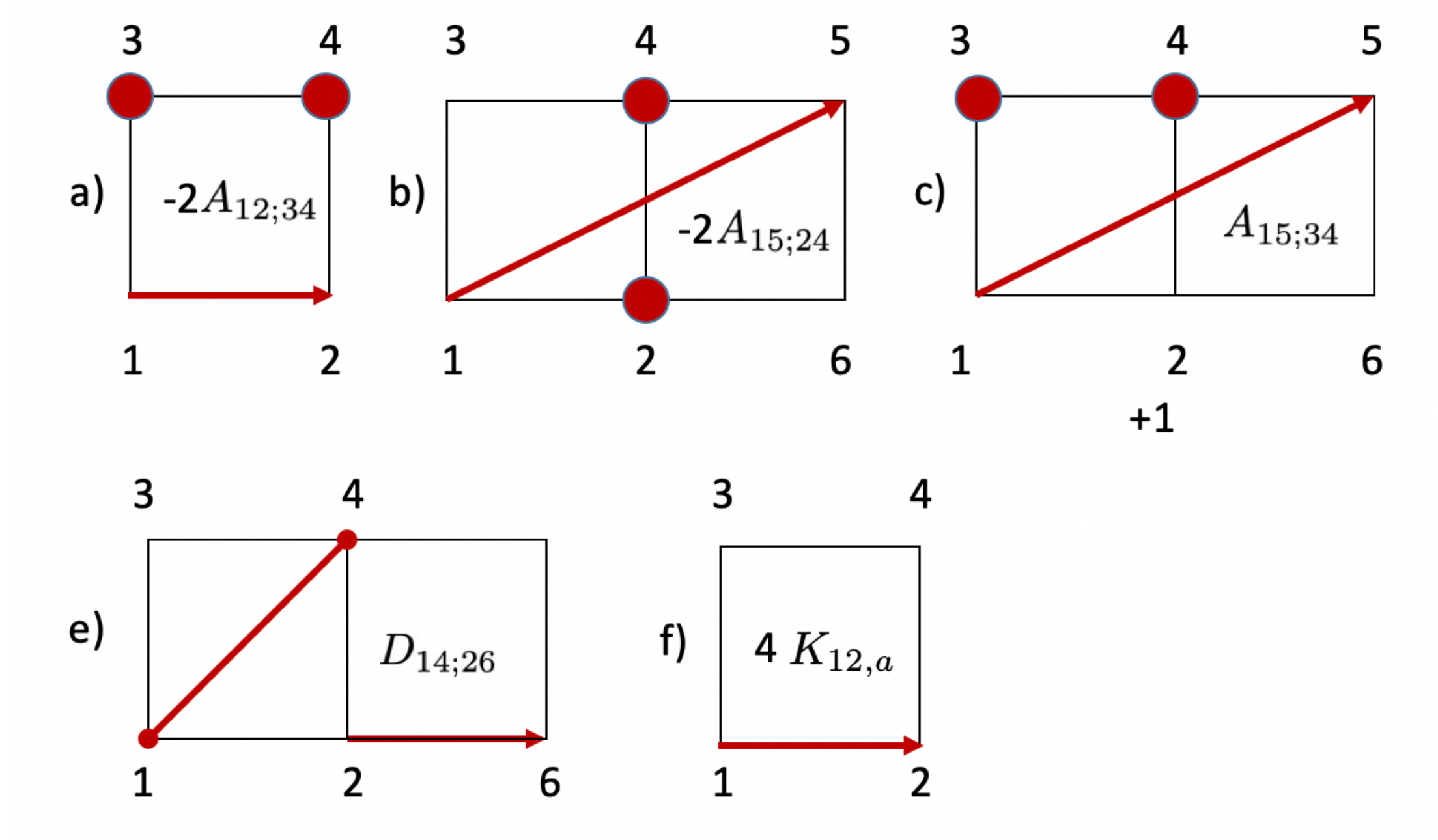}
\caption{Classes of operators of  $ \cL^2 j^x$, $K$, $A$ and $D$ are defined in Eqs.~(\ref{K-def}) and (\ref{OpL2}).}
 \label{fig:Delta2}
\end{center}
\end{figure}

 The  classes of operators 
are listed in Fig.~\ref{fig:Delta2}. The arrows mark the charge and spin bond operators and the circles mark 
  density and spin site operators. 
In the numbers $ a (b)$, $a$ is the lattice symmetry factor, and $(b)$ is the number of identical operators created by $(1-|0_x\rangle\langle 0_x|)\cL^2 \sum_i j^x_i$. 

The operators are,
\bea
A_{ij;kl} &=&  K_{ij;a}\left( {1\over 4} (1+n_k^h) (1+n_l^h) + s^\alpha_k s^\alpha_l\right)  \nonumber\\
&&+   \Sigma^\alpha_{ij;a} \left( s^\alpha_k  {1+n_l^h\over 2}  +s^\alpha_l  {1+n_k^h\over 2} \right) \nonumber\\
&&+ \epsilon_{\alpha\beta\gamma}\Sigma_{ij,s}^\alpha s_k^\beta s_l^\gamma \nonumber\\
D_{ij;kl} &=& {1\over 2} K_{ij;s}  \vec{K}_{jl }  + {1\over 2} \Sigma^\alpha_{ij;s}  \vec{\Sigma}^\alpha_{jl } .
\label{OpL2}
\eea     

\begin{figure}[h]
\begin{center}
\includegraphics[width=8cm,angle=0]{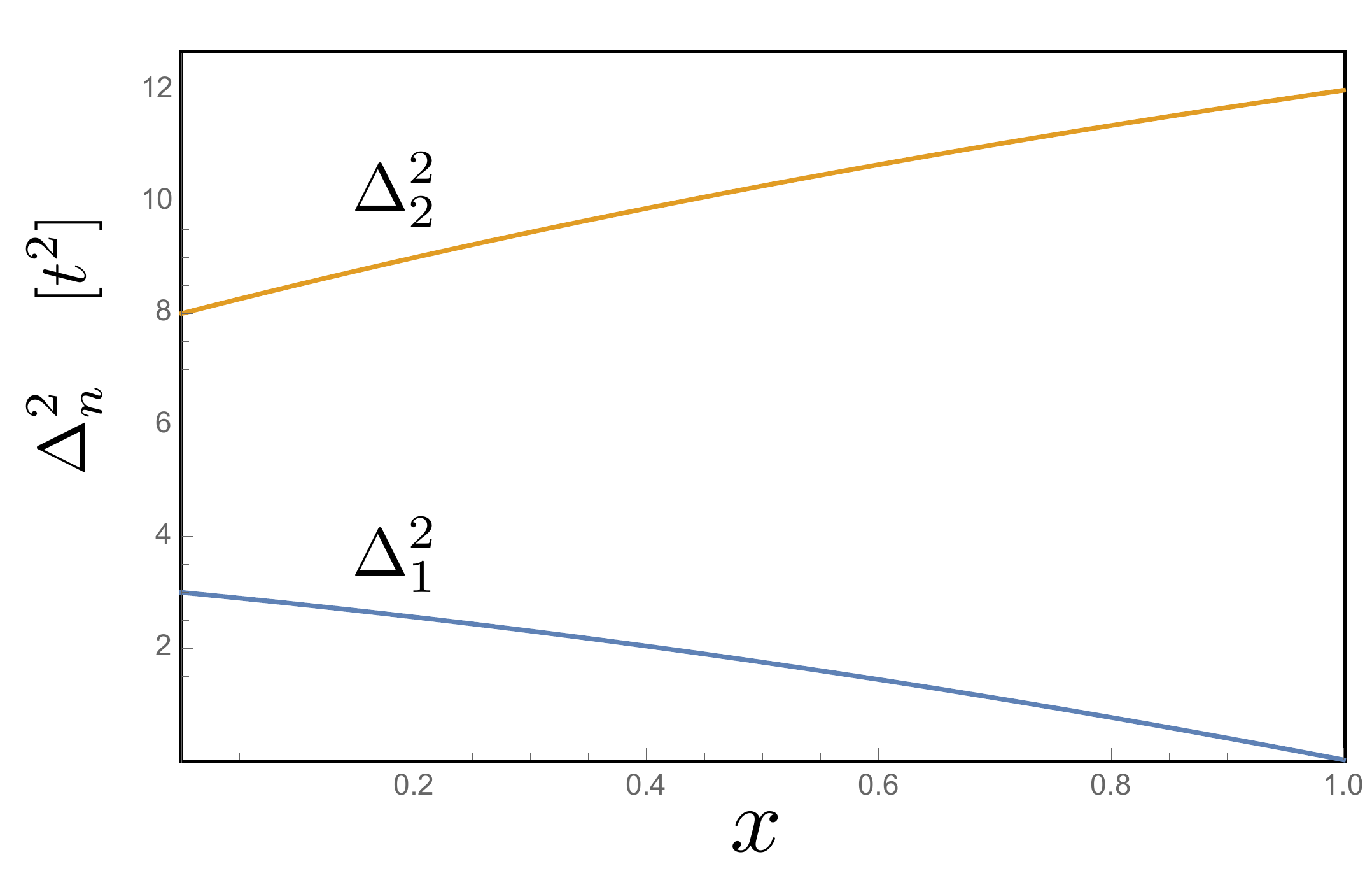}
\caption{The doping dependence of the first two recurrents of $H^t$ at high temperature.}
 \label{fig:DeltaRatios}
\end{center}
\end{figure}

 Thus,
 \bea
{\mu_4 \over \beta t^6}&=&  \overbrace{  \Tr (\underbrace{4}_{degen.}K_{12,a}^\dagger)(4 K_{12,a})}^{\mbox{bare curr}^2}       \nonumber\\
&&- 2\Tr (4K_{12,a}^\dagger) (2A_{12;34} + 2A_{12;78})      \nonumber\\
&&+  \Tr (2 A_{12;34}+2A_{12;78})^\dagger ( 2A_{12;34}+2A_{12;78})     \nonumber\\
&& +\underbrace{8}_{rotations} \Tr \left( A^\dagger_{15;34} A_{15;34} + A_{15;24} (2A_{15;24})\right.\nonumber\\
&&\left.  -  A^\dagger_{15;34} (2A_{15;24}) + A^\dagger_{15;34}A_{15;26}  \right) + 8\Tr D^\dagger D ,\nonumber\\
\eea
which yields 
\be
\Delta_2^2= { \mu_4 \over \Delta_1^2 \chi_{\rm csr} } -\Delta_1^2  = t^2 {24(1+x )\over 3 +x  }   \to_{x \to 0}  8 \quad  \to_{x \to 1} 12,
\ee
as depicted in Fig.~\ref{fig:DeltaRatios} and shown Eq.~(\ref{recurrents}). 

\section{Numerical calculation of recurrents and hypermagnetization matrix elements}
The moments $\mu_{2},\mu_4$ which yield the recurrents $\Delta_1,\Delta_2$ as well as the three hypermagnetization corrections $M''_{02},M''_{20},M''_{22}$  were evaluated numerically. 
They are written as expectation values of connected clusters of site operators. The clusters are  formed by commuting bond operators of the Hamiltonian or magnetization 
with the root current operator $j_i^\alpha$ on a single bond at site $i$. The result of $\cL^n j^\alpha_i$ is a large sum of  multi-site products of operators $O_{i_1}({\br_1}) \cdot  O_{i_2}(\br_2)\cdot ... \cdot O_{i_N}(\br_N)$, which is viewed as a product ``hyperstate''  in operator space,  with a complex amplitude that is stored separately.
Each application of the Liouvillian or the hypermagnetization can create a new hyperstate by  multiplying the individual site operators site-by-site using the multiplication Table~\ref{table:opmult}. 
One must  keep track of the order of the fermionic operators $\tilde{c}_i,\tilde{c}_j^\dagger$, and the negative signs produced when collecting contributions to the same product state from different multiplication paths.

\begin{table}[h!]
	\centering
	\begin{tabular}{ |c|c|c|c|c|c|c|c|c| } 
		\hline
		& $\tilde{c}^\dagger_\uparrow$ & $\tilde{c}^\dagger_\downarrow$ & $\tilde{c}_\uparrow$ & $\tilde{c}_\downarrow$ & $n_\uparrow$ & $n_\downarrow$ & $ \tilde{s}^+$ & $\tilde{s}^-$ \\ 
		\hline
		$\tilde{c}^\dagger_\uparrow$ & 0 & 0 & $n_\uparrow$ & $\tilde{s}^+$ & 0 & 0 & 0 & 0 \\ 
		\hline
		$\tilde{c}^\dagger_\downarrow$ & 0 & 0 & $\tilde{s}^-$ & $n_\downarrow$ & 0 & 0 & 0 & 0 \\ 
		\hline
		$\tilde{c}_\uparrow$ & $n_h$ & 0 & 0 & 0 & $\tilde{c}_\uparrow$ & 0 & $\tilde{c}_\downarrow$ & 0 \\ 
		\hline
		$\tilde{c}_\downarrow$ & 0 & $n_h$ & 0 & 0 & 0 & $\tilde{c}_\downarrow$ & 0 & $\tilde{c}_\uparrow$ \\ 
		\hline
		$n_\uparrow$ & $\tilde{c}^\dagger_\uparrow$ & 0 & 0 & 0 & $n_\uparrow$ & 0 & $\tilde{s}^+$ & 0 \\ 
		\hline
		$n_\downarrow$ & 0 & $\tilde{c}^\dagger_\downarrow$ & 0 & 0 & 0 & $n_\downarrow$ & 0 & $\tilde{s}^-$ \\ 
		\hline
		$\tilde{s}^+$ & 0 & $\tilde{c}^\dagger_\uparrow$ & 0 & 0 & 0 & $\tilde{s}^+$ & 0 & $n_\uparrow$ \\ 
		\hline
		$\tilde{s}^-$ & $\tilde{c}^\dagger_\downarrow$ & 0 & 0 & 0 & $\tilde{s}^-$ & 0 & $n_\downarrow$ & 0  \\ 
		\hline
	\end{tabular}\caption{Multiplication table of GP operators in the tJM. The entry $O_{i,j}=O_i \cdot  O_j$, where $i$ and $j$ are row and column respectively.}
	\label{table:opmult}
\end{table}
where,
\be
\tilde{s}^\alpha=s^\alpha (1-n^h),\quad n_{\uparrow,\downarrow}= (1-n^h)\left({1\over 2}\pm s^z\right).
\ee
For traces however, most operators vanish unless they contain only constants and factors of density $n_\gamma$, (see Eqs.~(\ref{Tr1}-\ref{K3}).
\section{QMC sign errors}
\label{App:QMCsign}
  \begin{figure}[h]
\includegraphics[width=8cm,angle=0]{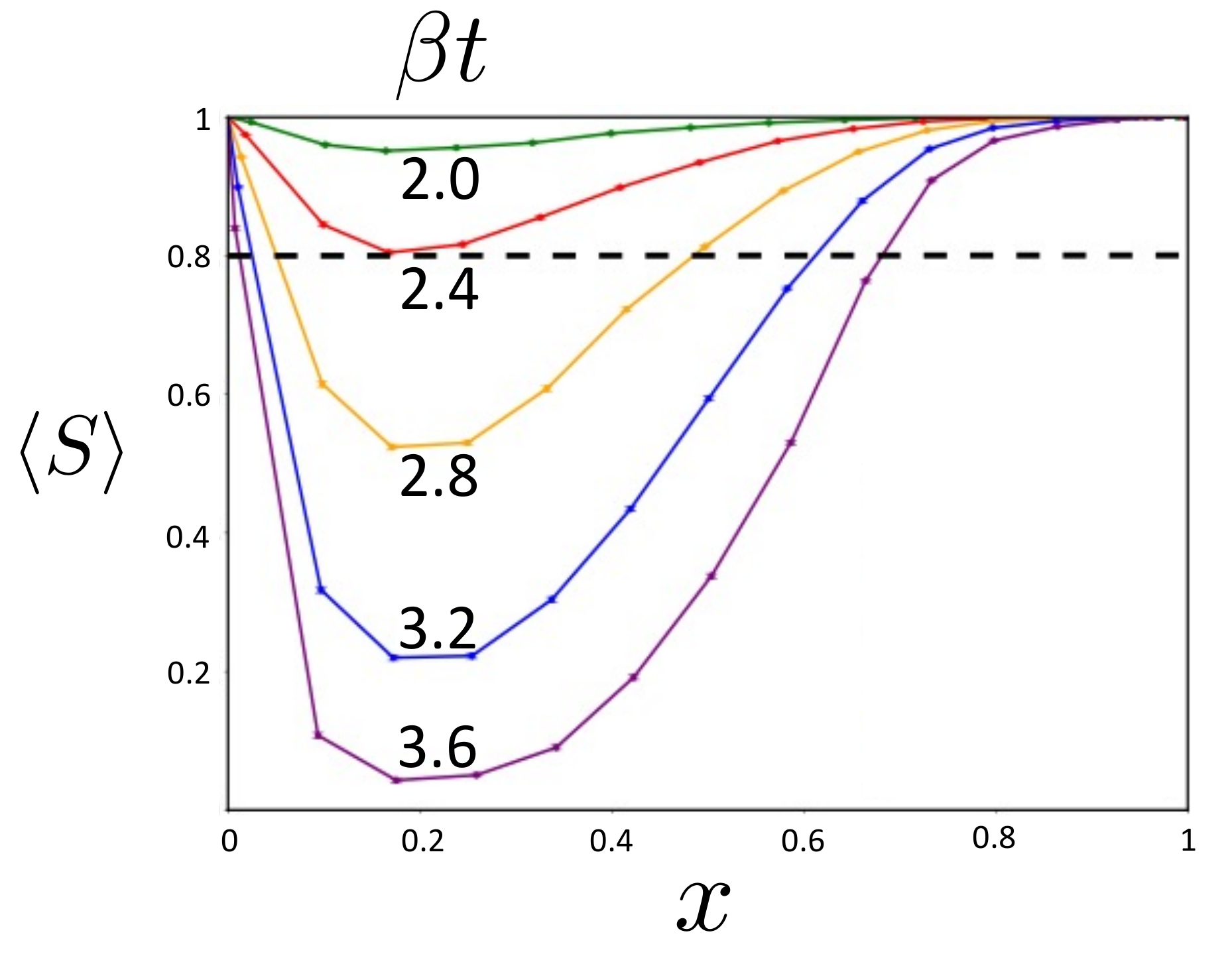}
\caption{QMC fermion sign average for the HM  as a function of doping for $U=8t$, and for a set of inverse temperatures $\beta$. }   \label{fig:Sign-vsT} 
\end{figure}
  \begin{figure}[h]
\includegraphics[width=8cm,angle=0]{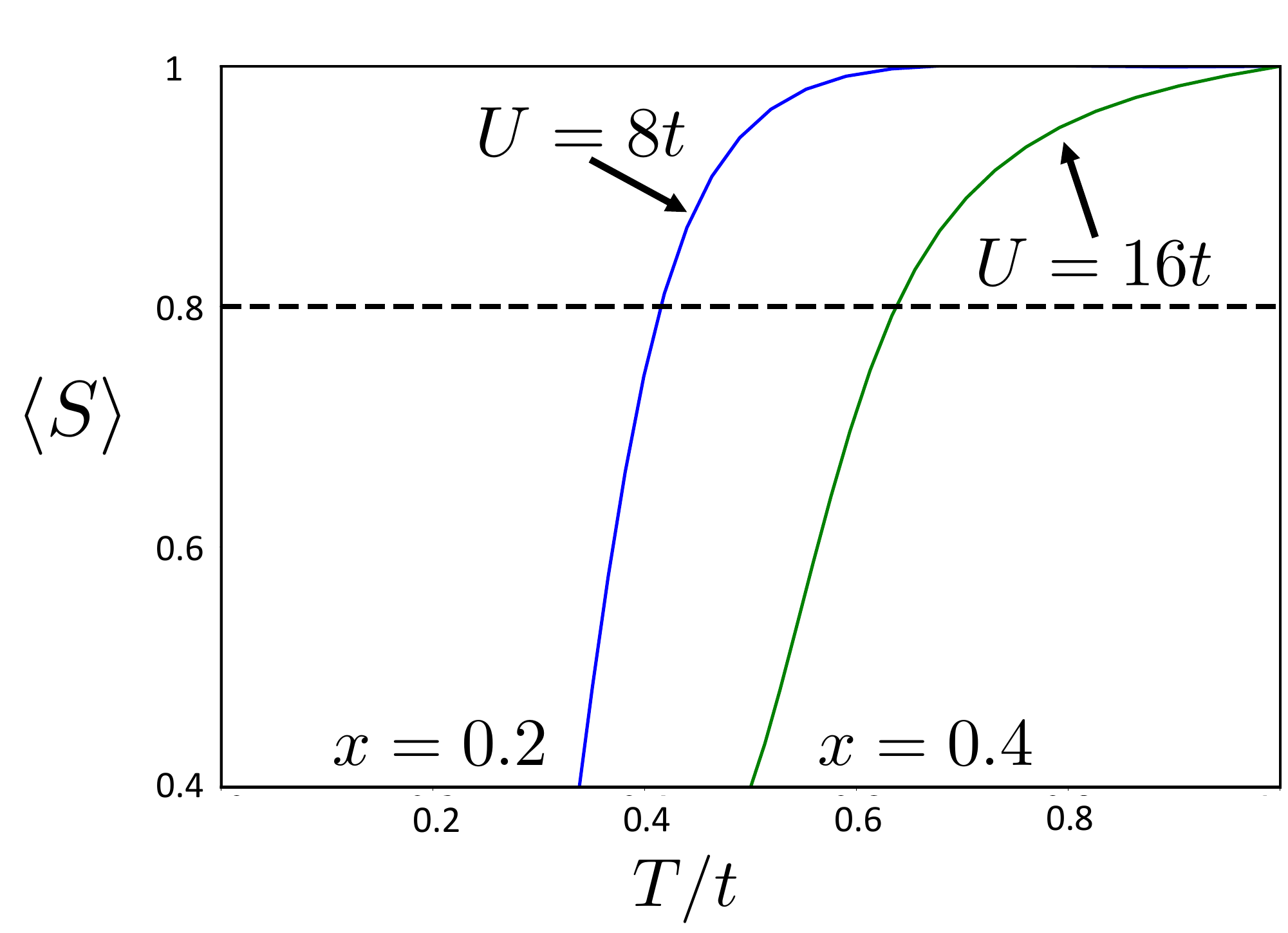}
\caption{QMC fermion sign average  for the HM, depicted as a function  of  temperature for  $U=8t$ at $x=0.2$,  and  $U=16t$ at $x=0.4$. }   \label{fig:Sign-vsU} 
\end{figure}

The QMC method becomes unstable at low temperatures, as is well known, 
due to the fermion {\em sign problem}~\cite{Sign}. The determinant emerging 
from integrating out fermions can be negative for certain configurations, and consequently unsuitable  
as a sampling weight. This effect is quantified by plotting the 
average sign as a function of doping and temperature, as shown in Fig.~\ref{fig:Sign-vsT}. It is revealed that for $U=8t$, one may reduce $T$ down to $ 0.4 t$, before
the average sign $\langle S\rangle=\langle \sgn({\rm det})\rangle$ falls below 0.8. 
 We have not utilized the QMC below this temperature.
In  Fig.~\ref{fig:Sign-vsU} we plot $\langle S\rangle$ as a function of temperature for two 
interaction strengths $U=8t,16t$ and choosing the ``worst''  doping levels for these interactions, at $x=0.2$
and $x=0.4$, respectively. Here  we observe that the  ``sign error 
temperature''   increases with interaction strength.

\end{document}